\documentclass[sigconf]{acmart}
\AtBeginDocument{%
  }

\setcopyright{acmlicensed}
\copyrightyear{2018}
\acmYear{2018}
\acmDOI{XXXXXXX.XXXXXXX}
\acmConference[Conference acronym 'XX]{Make sure to enter the correct
  conference title from your rights confirmation email}{June 03--05,
  2018}{Woodstock, NY}
\acmISBN{978-1-4503-XXXX-X/2018/06}

\usepackage{multirow}
\usepackage{booktabs}   % 用于绘制漂亮的表格横线
\usepackage{makecell}   % 用于单元格内换行
\usepackage{graphicx}   % 用于 \resizebox 调整表格大小
\usepackage{subfig}
\usepackage{enumitem}
\usepackage{booktabs} % 用于漂亮的表格横线
\usepackage{ulem}
\usepackage{xcolor}
\usepackage{hyperref}
\begin{document}

\title{WeaveRec: An LLM-Based Cross-Domain Sequential Recommendation Framework with Model Merging}

\author{Min Hou}
\email{hmhoumin@gmail.com}
\affiliation{%
  \institution{Hefei University of Technology}
  \city{Hefei}
  \country{China}
}

\author{Xin Liu}
\email{xinliu221b@gmail.com}
\affiliation{%
  \institution{Hefei University of Technology}
  \city{Hefei}
  \country{China}
}

\author{Le Wu*}\thanks{*Corresponding Author}
\email{lewu.ustc@gmail.com}
\affiliation{%
  \institution{Hefei University of Technology}
  \city{Hefei}
  \country{China}
}

\author{Chenyi He}
\email{hechenyi@mail.hfut.edu.cn}
\affiliation{%
  \institution{Hefei University of Technology}
  \city{Hefei}
  \country{China}
}

\author{Hao Liu}
\email{haoliu@mail.hfut.edu.cn}
\affiliation{%
 \institution{Hefei University of Technology}
 \city{Hefei}
 \country{China}}

\author{Zhi Li}
\email{zhilizl@sz.tsinghua.edu.cn}
\affiliation{%
  \institution{Tsinghua University}
  \city{Shenzhen}
  \country{China}}

\author{Xin Li}
\affiliation{%
  \institution{iFLYTEK}
  \city{Hefei}
  \country{China}}
\email{	leexin@ustc.edu.cn}

\author{Si Wei}
\affiliation{%
  \institution{iFLYTEK}
  \city{Hefei}
  \country{China}}
\email{siwei@iflytek.com}

\renewcommand{\shortauthors}{Trovato et al.}

\begin{abstract}
Cross-Domain Sequential Recommendation~(CDSR) seeks to improve user preference modeling by transferring knowledge from multiple domains.
Despite the progress made in CDSR, most existing methods rely on overlapping users or items to establish cross-domain correlations-a requirement that rarely holds in real-world settings.
The advent of large language models~(LLM) and model-merging techniques
appears to overcome this limitation by unifying multi-domain data without explicit overlaps. Yet, our empirical study shows that naively training an LLM on combined domains—or simply merging several domain‐specific LLMs—often degrades performance relative to a model trained solely on the target domain.

To address these challenges, we first experimentally investigate the cause of suboptimal performance in LLM-based cross-domain recommendation and model merging. Building on these insights, we introduce WeaveRec, which cross-trains multiple LoRA modules with source and target domain data in a "weaving" fashion, and fuses them via model merging. WeaveRec can be extended to multi-source domain scenarios and notably does not introduce additional inference-time cost in terms of latency or memory.
Furthermore, we provide a theoretical guarantee that WeaveRec can reduce the upper bound of the expected error in the target domain.
Extensive experiments on single-source, multi-source, and cross-platform cross-domain recommendation scenarios validate that WeaveRec effectively mitigates performance degradation and consistently outperforms baseline approaches in real‐world recommendation tasks. Codes are available at \url{https://anonymous.4open.science/r/WeaveRec-829F}.

\iffalse
Large Language Models (LLMs) show significant potential in empowering recommender systems. Nevertheless, recommender systems still face the challenge of data sparsity. Cross-domain recommendation aims to alleviate this by introducing knowledge from one or more source domains to enhance model performance in a target domain. But the majority of existing CDR methods rely on the assumption of overlapping users between domains. Some efforts address non-overlapping scenarios, they typically approach the problem from the perspective of optimizing embedding vectors. From a novel model merging perspective, We proposes \textit{WeaveRec}, a low-cost and effective LoRA merging framework for LLM-based cross-domain sequential recommendation. \textit{WeaveRec} utilizes one or more source domain LoRAs, effectively enhancing the LLM's recommendation knowledge for the target domain and boosting its performance, regardless of user overlap between different domains. Extensive experimental results demonstrate that \textit{WeaveRec} effectively mitigates the negative transfer problem, successfully achieving cross-domain recommendation through model merging.
\fi
\end{abstract}



\keywords{Do, Not, Us, This, Code, Put, the, Correct, Terms, for,
  Your, Paper}

\received{20 February 2007}
\received[revised]{12 March 2009}
\received[accepted]{5 June 2009}

\maketitle

\section{Introduction} \label{intro}
\begin{figure}[!t]
    \centering
    \includegraphics[width=0.48\textwidth]{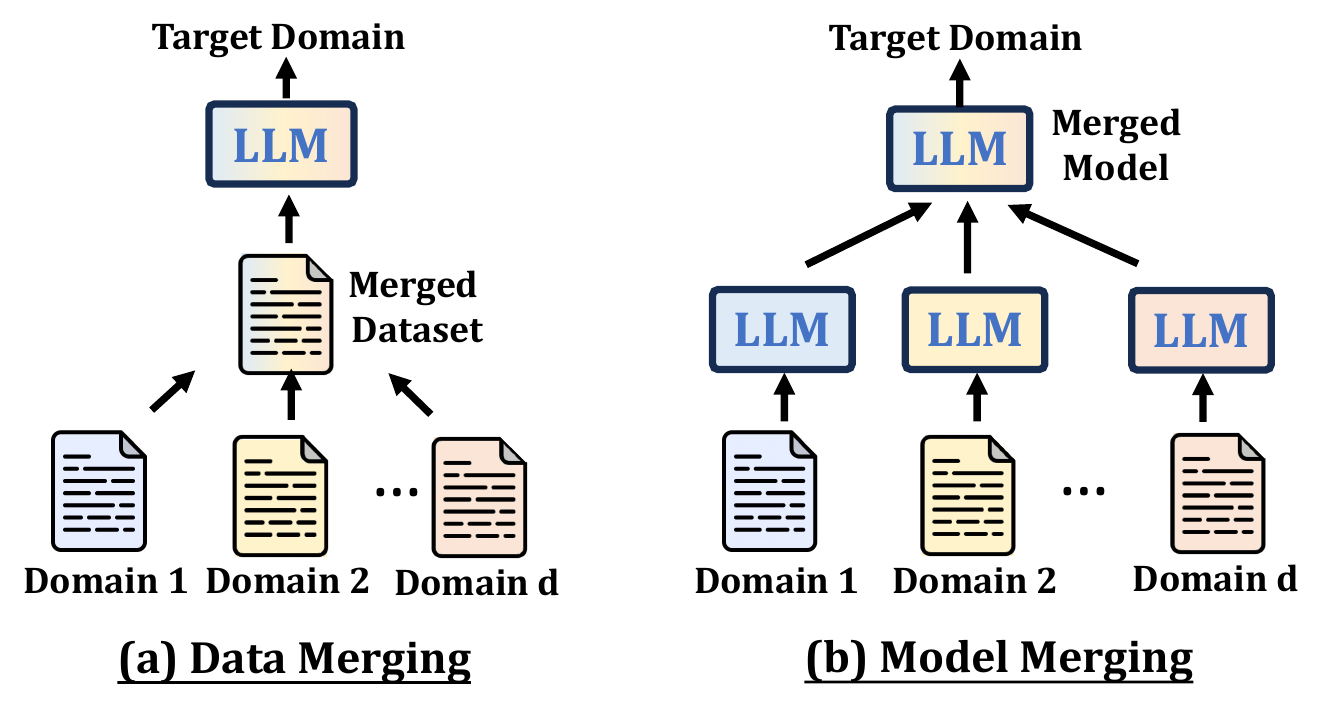}
    \caption{Illustration of Data Merging and Model Merging.}
    \label{fig:intro_1}
    \Description{}
\end{figure}
With the rapid growth of the Internet, a wide range of online services has emerged, generating vast volumes of user interactions across multiple domains. Each domain encodes valuable behavioral signals and preference data. Cross‑domain sequential recommendation (CDSR)~\cite{10.1145/3548455,ijcai2021p639,zhang2025comprehensivesurveycrossdomainrecommendation} has therefore arisen as a powerful approach, leveraging knowledge transfer from source domains to bolster recommendation performance in a target domain, addressing the fundamental challenge of data sparsity in individual domains. Along this line, existing traditional CDSR methodologies can be divided into two primary categories based on their representation strategies. 1) ID-based approaches~\cite{10.1109/TKDE.2022.3185101,10.1609/aaai.v38i8.28723,10.1145/3539597.3570366,10.1109/TKDE.2024.3511602} 
employ collaborative filtering models to learn domain-specific embeddings, which are subsequently aligned through overlapping users or items via techniques such as mapping functions or shared latent spaces. 
2) Transferable approaches~\cite{unisrec,recformer,hou2023vqrec} employ content-based representations, particularly textual descriptions, to encode items within a unified semantic space, enabling the learning of universal and transferable sequence representations across domains.

\iffalse
While effective when overlap exists, these methods face severe scalability constraints due to their dependency on cross-domain overlaps, which are often sparse or unavailable in practice.
Transferable approaches~\cite{unisrec,recformer,hou2023vqrec} address this limitation by utilizing content-based representations, particularly textual descriptions, to encode items in a unified semantic space. Notable examples include VQ-Rec~\cite{hou2023vqrec} and UniSRec~\cite{unisrec}, which employ contrastive pre-training on language models to learn domain-agnostic representations that facilitate knowledge transfer without requiring explicit overlaps.
\fi

\begin{figure} [!t]
    \centering
    \subfloat[Target domain: Clothing]{\includegraphics[width=0.48\columnwidth]{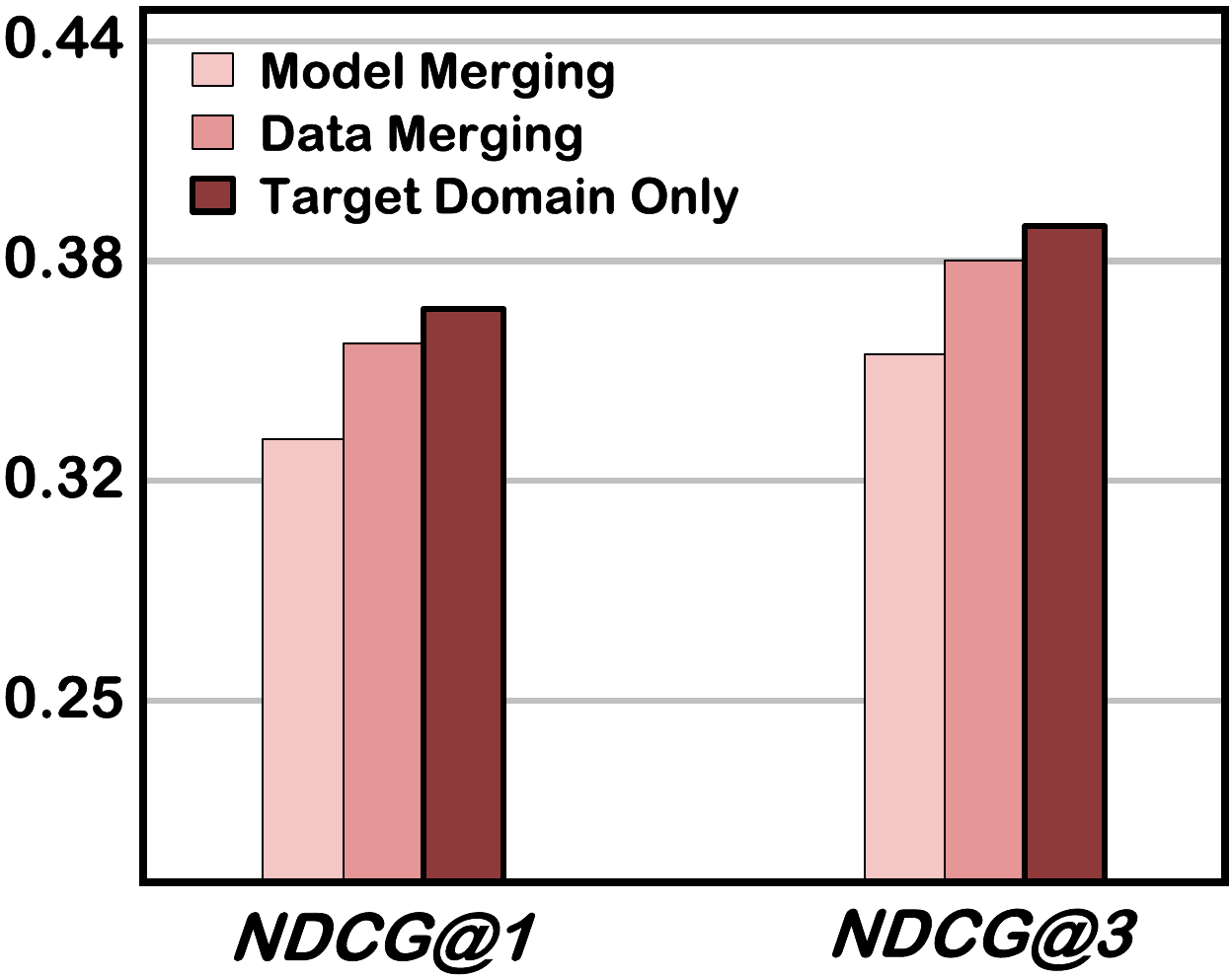}}
    \hfil
    \subfloat[Target domain: Beauty]{\includegraphics[width=0.48\columnwidth]{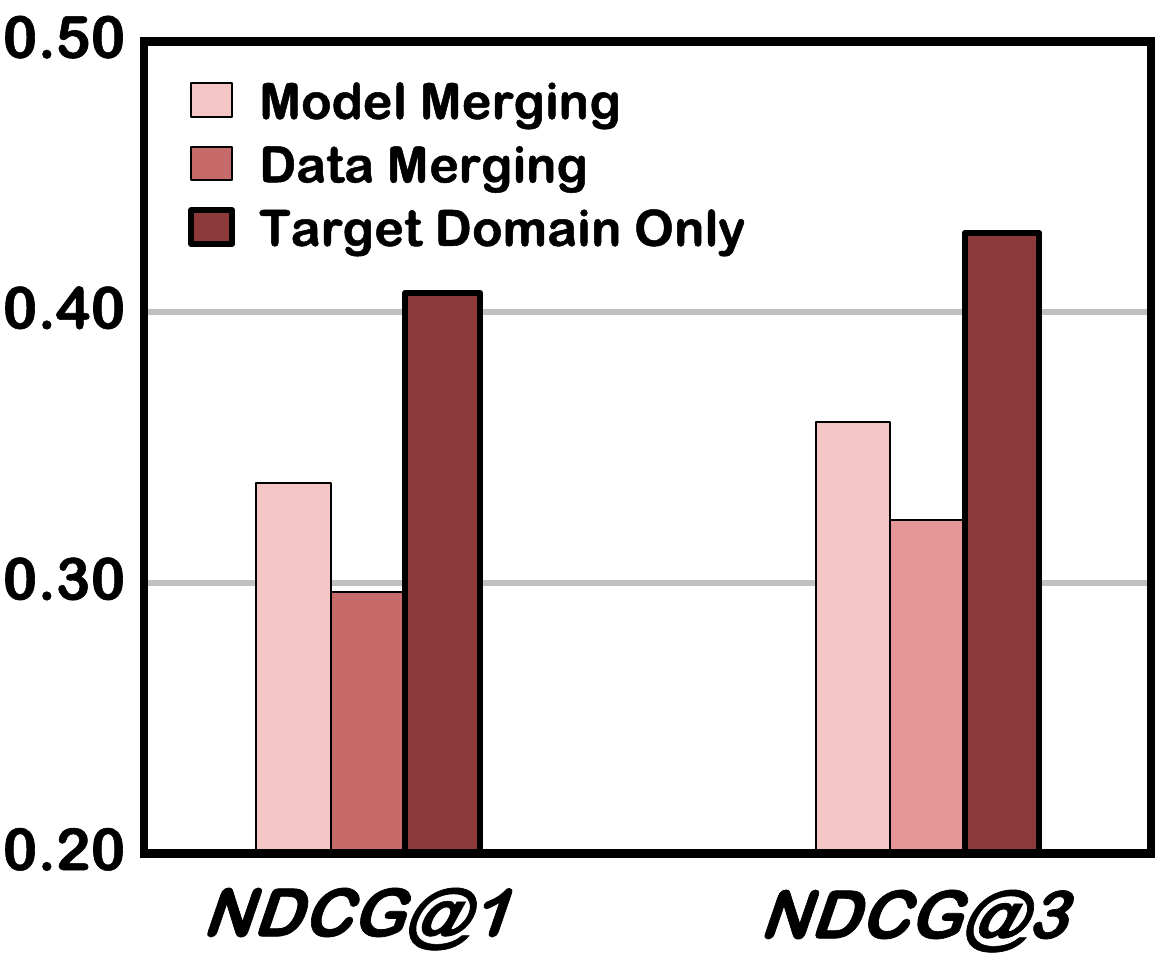}}
    \caption{Illustration of performance degradation under data merging and model merging. Experiments are based on four source domains: Amazon Beauty, Sports, Clothing, and Food, and two target domains: Clothing and Beauty.}
    \label{fig:intro-chart}
\end{figure}

Recently, Large Language Models (LLMs) have demonstrated remarkable success across diverse fields~\cite{zhao2023survey}, driven by their emergent capabilities~\cite{dong2024survey,huang-chang-2023-towards} such as world knowledge, language understanding, and complex reasoning. Building upon these strengths, LLMs shift recommender systems from task-specific designs to unified, general-purpose models capable of handling diverse domains and tasks~\cite{LLM4CDR,one-model-for-all,p5,m6rec,ecellm}, and further introduced transformative advancements to CDSR~\cite{one-model-for-all,10.1145/3523227.3546767,10.5555/3692070.3693702}. The core methodology involves aggregating multi-domain and multi-task recommendation data into unified instruction-tuning datasets, followed by training a single comprehensive model capable of handling diverse domains and tasks~\cite{LLM4CDR,one-model-for-all,p5,m6rec,ecellm}. This "one model for all" paradigm effectively overcomes traditional CDSR constraints such as dependency on overlapping users/items and limited representation capabilities.
Representative works include M6-rec~\cite{m6rec}, which develops a foundation model supporting open-ended domains and tasks in industrial settings; LLM-Rec~\cite{one-model-for-all}, which explores language models' capabilities in modeling multi-domain user behavior. However, we argue that this approach of aggregating multi-source data to directly train a model has the following limitations: \textbf{1) Inflexible}. The addition or removal of a domain necessitates model retraining from scratch,  resulting in prohibitive computational costs and limited practical applicability.\textbf{ 2) Data Conflict.} User interactions from different domains often contain conflict (i.e., interactions irrelevant to or conflicting with the target domain’s recommendation), which
leads to model misaligned with users’ true preferences in the target domain and ultimately degrades recommendation performance. Our preliminary empirical analysis in Figure \ref{fig:intro-chart} reveals a critical phenomenon: data merging often yields performance degradation compared to target-domain-only models. These limitations motivate us to find a new paradigm for building a unified CDSR model.

Fortunately, model merging~\cite{Survery_ModelMerging_2024} offers a viable alternative by combining model parameters in weight, as shown in Figure \ref{fig:intro_1}(b). By merging multiple single-task models' parameters, model merging is designed to obtain a unified model that can simultaneously perform multiple tasks without the need for retraining. This is an exciting and promising technology, which is being applied to various scenarios, such as unlearning old knowledge in LLMs\cite{fuseforget}, achieving image-style transformation\cite{diffusionsoup}, and so on.
If the model merging technique can be applied in CDSR, then it can naturally solve the limitations of inflexibility and data conflict.
This is because if we want to add or remove a source domain, we only need to operate on the saved model parameters without retraining models for all the other domains. Furthermore, each domain's model is trained using only its specific domain data, which reduces the impact of data conflicts. However, naively applying model merging to cross-domain recommendation presents significant challenges. Experiments in Figure \ref{fig:intro-chart} reveal that model merging still suffers from the performance degradation in the target domain. This degradation occurs when source domain knowledge conflicts with target domain patterns, causing the merged model to converge to suboptimal representations that satisfy neither domain effectively.

In this paper, we explore the integration of model‑merging techniques into cross‑domain recommendation. Our aim is to preserve the inherent scalability and extensibility of model merging while ensuring consistent performance improvements on the target domain. 
We first experimentally analyze potential causes of performance degradation in model merging techniques for LLM-based cross-domain recommendation. \uline{Experiments suggest that the performance degradation is more likely to occur when the source-domain model performs poorly in the target domain.} Specifically, when it happens, the source-domain models capture patterns that are not only irrelevant but actively misleading for the target domain recommendation. The poor source-domain model "drags" the fused network into a compromise that fits none of the domains well, manifesting as severe performance degradation on the target domain. Based on the findings above, our goal shifts to improving the performance of the source domain model on the target domain.
However, ensuring the source model's performance on the target domain is not trivial. The source and target domains often exhibit significant differences in user behavior patterns, item characteristics, and interaction distributions, making it difficult for source-domain models to generalize effectively to the target domain without substantial adaptation.
To solve this challenge, we present a simple but effective model merging-based cross-domain recommendation framework, named \textit{WeaveRec}. We train a model using mixed data from the source domain and the target domain and merge it with the target-domain-only model.
In such a "weave"-like manner, the new source domain model can be better adapted to the target domain distribution, therefore avoiding performance degradation in the target domain. We also extend WeaveRec to multi-source domain scenarios and notably do not introduce additional inference-time cost in terms of latency or memory. Furthermore, our theoretical analysis demonstrates that WeaveRec effectively ensure the source domain model's performance on the target domain by provably reducing the upper bound of generalization error in the target domain. Extensive experiments on single-source, multi-source, and cross-platform cross-domain recommendation scenarios validate that WeaveRec consistently outperforms baseline approaches in real-world recommendation tasks. The main contributions of this work are as follows:
\begin{itemize}[leftmargin=*]
    \item We propose WeaveRec, a simple yet effective framework that demonstrably stable performance improvement while maintaining the scalability advantages of model merging. 
    \item We provide an analysis of performance degradation in model merging for CDSR. 
    Our theoretical analysis demonstrates that WeaveRec effectively ensure the source domain model's performance on the target domain by provably reducing the upper bound of generalization error in the target domain.
    \item Extensive experiments on single-source, multi-source, and cross-platform cross-domain recommendation scenarios validate the effectiveness of WeaveRec.
\end{itemize}

\section{Preliminaries} \label{Preliminaries}
\noindent$\bullet$ \quad\textbf{CDSR Task Formulation.}
Cross-Domain Sequential Recommendation (CDSR) aims to predict users’ preferences based on historical sequential interactions across multiple domains. Formally, we denote the set of domains as $\mathcal{D}=\{D_0, D_1, ..., D_N\}$ where $D_0$ denotes the target domain, $\{D_n\}_{n=1}^{N}$ denotes at least one source domain. Thus the number of source domains $N\geq1$, and $|\mathcal{D}|\geq2$. We define $\mathcal{S}_n$, $\mathcal{U}_n$ and $\mathcal{V}_n$ as the set of user interaction sequences, users, and items, respectively, in the domain $D_n$, $0\leq n\leq N$. In an arbitrary domain, the interaction sequences of users are ordered chronologically. For example, let $u\in\mathcal{U}_n$ be a particular user in domain $D_n$, the sequence $s_u\in\mathcal{S}_n$ can be represented by $[v_1, v_2,...,v_{|s_u|}]$, where the subscript denotes the time step and all the items of the sequence belong to domain $D_n$.
The goal of CDSR is to predict the next most likely item $v_{|s_u|+1}$ for \textbf{users in the target domain}, based on their historical sequences. Formally, this goal can be expressed as:
\begin{equation}
max\ P\{v_{|s_u|+1}=v\ |\ s_u,\mathcal{K}(\{D_n\}_{n=0}^N)\},\  \forall u\in \mathcal{U}_0,
\end{equation}
where $v$ refers to the corresponding ground truth and $\mathcal{K}(\cdot)$ represents the knowledge learned from all domains.

\noindent$\bullet$ \quad\textbf{Instruction Tuning for LLM-Based Recommendation.} For LLM-based, instruction tuning is the key step to bridge the gap between the general task of next-word prediction and the recommendation task. Specifically, we need to prepare explicit instruction pairs $\{(\textbf{x}_u, \textbf{y}_u)\mid u\in \mathcal{U}\}$, where $\textbf{x}_u$ stands for a specific textual input that includes the user's historical sequence and a candidate set, and $\textbf{y}_u$ is the label which contains text (e.g. title or other descriptions) of the real next item. The fine-tuning is guided by minimizing the negative log-likelihood loss function:
\begin{equation}
\Theta^*=argmin_{\Theta}\{-\sum_u\sum_{t=1}^{|\textbf{y}_u|}logP_{\Theta}(y_u^t\mid \textbf{y}_u^{<t}, \textbf{x}_u)\}, \label{eq1}
\end{equation}
where $\Theta$ denotes LLM's parameters, $y_u^t$ indicates the $t$-th token of $\textbf{y}_u$ and $\textbf{y}_u^{<t}$ is the token sequence from the previous $t$ time steps.

Duo to the immense size of LLMs, the cost of updating all parameters is prohibitively expensive. Consequently, Parameter-Efficient Fine-Tuning (PEFT) emerged, which adjusts a small part of parameters while keeping most of the pre-trained model's parameters frozen. LoRA\cite{2022lora} is one of the representative PEFT techniques. LoRA adapts LLMs to a new task by introducing low-rank matrices into the model's linear layers, without altering the model's original parameters. 
Specifically, for any pre-trained weight matrix $\textbf{W}\in\mathbb{R}^{d_{out}\times d_{in}}$ in the transformer block of the LLM, which takes an input vector $\boldsymbol{x}\in\mathbb{R}^{d_{in}}$ and outputs $\boldsymbol{h}\in\mathbb{R}^{d_{out}}$. LoRA changes $\boldsymbol{h=Wx}$ to:
\begin{equation}
\boldsymbol{h=Wx+BAx}, \label{eq2}    
\end{equation}
where $\boldsymbol{B} \in\mathbb{R}^{d_{out}\times r}$, $\boldsymbol{A} \in\mathbb{R}^{r\times d_{in}}$ are low-rank projection matrices. It is worth noting that the rank $r\ll min(d_{in},d_{out})$, meaning that the number of trainable parameters introduced by $\boldsymbol{BA}$ is significantly less than those of $\boldsymbol{W}$. During fine-tuning with LoRA, The LLM's own parameters are frozen, and only the $\boldsymbol{BA}$ matrices are updated. Here we denote $\theta$ as additional parameters introduce by LoRA. Therefore, Eqn. (\ref{eq1}) can be rewritten as:
\begin{equation}
    \theta^*=argmin_{\theta}\{-\sum_u\sum_{t=1}^{|\textbf{y}_u|}logP_{\Theta+\theta}(y_u^t\mid \textbf{y}_u^{<t}, \textbf{x}_u)\}, \label{eq3}
\end{equation}
where $\theta=\{\boldsymbol{B}^l,\boldsymbol{A}^l\}_{l=1}^L$ denotes the set of initialized LoRA parameters, and $L$ is the number of LoRA modules.

\noindent$\bullet$ \quad\textbf{Data Merging for LLM-Based CDSR.} The emergence of LLMs has enabled a paradigm shift in recommender systems from task-specific architectures to unified, general-purpose models capable of handling diverse domains and tasks. A prevalent approach involves consolidating recommendation data from both source and target domains into a unified instruction-tuning dataset, followed by supervised fine-tuning of pre-trained LLM backbones on it. The resulting model encapsulates knowledge from multiple domains, enabling recommendation to be performed on the target domain.

\noindent$\bullet$ \quad\textbf{Naive Model Merging for LLM-Based CDSR.}
Model merging is rooted in the theoretical foundation of mode connectivity~\cite{1996Weight,10.5555/3327546.3327556,frankle2020linear}, \uline{the principle that models fine-tuned from the same pre-trained checkpoint often reside in connected regions of the loss landscape, enabling meaningful parameter interpolation without significant performance degradation.}
The principle enables us to train multiple LoRAs for each domain separately, and then merge them together. For the CDSR task, formally, given the recommendation data from multiple domains $\mathcal{D}=\{D_0, D_1, ..., D_N\}$, based on Eqn. (\ref{eq3}), we can train one LoRA module $\theta_n=\{\boldsymbol{B}_n^l,\boldsymbol{A}_n^l\}_{l=1}^L$ for each domain $n$. Then we merge the LoRA models through weight averaging: 
\begin{equation}
    \theta_m = (\frac{1}{N+1}\theta_0) \oplus (\frac{1}{N+1}\theta_1) \oplus \cdot \cdot \cdot \oplus (\frac{1}{N+1}\theta_N) = \{\boldsymbol{A}_m^l, \boldsymbol{B}_m^l\}_{l=1}^L,
\label{eq:merge}
\end{equation}
\begin{equation}
    \boldsymbol{A}_m^l = \frac{1}{N+1}\boldsymbol{A}_0^l + \frac{1}{N+1}\boldsymbol{A}_1^l + \cdot\cdot\cdot + \frac{1}{N+1}\boldsymbol{A}_N^l,
\end{equation}
\begin{equation}
    \boldsymbol{B}_m^l = \frac{1}{N+1}\boldsymbol{B}_0^l + \frac{1}{N+1}\boldsymbol{B}_1^l + \cdot\cdot\cdot + \frac{1}{N+1}\boldsymbol{B}_N^l,
\end{equation}
The merged LoRA $\theta_m$ maintains the same total number of parameters as one standard LoRA. In addition, the LoRA module is reusable. It is easy to remove or add knowledge from a specific domain without retraining the whole model. Although this naive model merging approach possesses attractive properties and is widely used in multi-task learning scenarios, it cannot be directly applied to CDSR tasks due to potential phenomena of performance degradation in the target domain.

\section{Experimental Analysis} \label{analysis}
Current research efforts have given rise to numerous model merging methods. This raises the question: What effect would applying these methods to cross-domain recommendation tasks have? To address this, we conduct experiments on CDSR tasks using existing model merging techniques and analyze the observations.

% \subsection{\textbf{Experimental Observations}} \label{last of preli}
\noindent$\bullet$ \quad\textbf{Experimental Settings.}
Some representative model merging methods are selected for experimentation. We choose Amazon Sports as the target domain and Clothing as the source domain. Thus, we can obtain two distinct LoRAs, which have learned recommendation knowledge from the two domains, respectively. They are then merged into a single new LoRA using the chosen methods, and its performance is evaluated on the target domain. The chosen methods are as follows.

\noindent$\bullet$ \quad\textbf{Model Merging Methods.}
\textbf{(1) Weight Average(WA)}~\cite{wortsman2022model} is the simplest model merging method, directly combining multiple single-task/domain models by their average weights, as described in Eqn. (\ref{eq:merge}). \textbf{(2) Ext-Sub} \cite{ext-sub} decomposes LoRA modules from different tasks into shared and task-specific components to mitigate inter-task conflicts during model merging.
\textbf{(3) DARE} \cite{DARE} mitigates parameter interference in model merging by eliminating a significant number of redundant parameters, and it can be integrated with any downstream model merging method. 
\textbf{(4) LoRA-LEGO} \cite{lego} is a LoRA merging technique, which decouples each LoRA into several Minimum Semantic Units (MSUs) and then clusters all of them to form a new merged LoRA. 
\textbf{(5) Tie-Merging} \cite{ties} involves a three-step process that includes reducing parameter redundancy, eliminating sign conflicts between parameters, and finally merging them. 
\textbf{(6) AdaMerging} \cite{AdaMerging_ICLR_2024} is an adaptive model merging technique that automatically learns optimal merging coefficients(rather than using uniform coefficients) for multi-task learning by leveraging entropy minimization on unlabeled test data.

% \iffalse
% \begin{figure} [htbp]
%     \centering
%     \includegraphics[width=0.6\linewidth]{figures/preliminary-comparison.pdf}
%     \caption{Performance of model merging methods compared to target-domain only on the target domain.}
%     \label{preliminary-comparison}
% \end{figure}
% \fi

\begin{figure}[htbp]
    \centering
    \subfloat[Clothing as source domain]{\includegraphics[width=0.49\columnwidth]{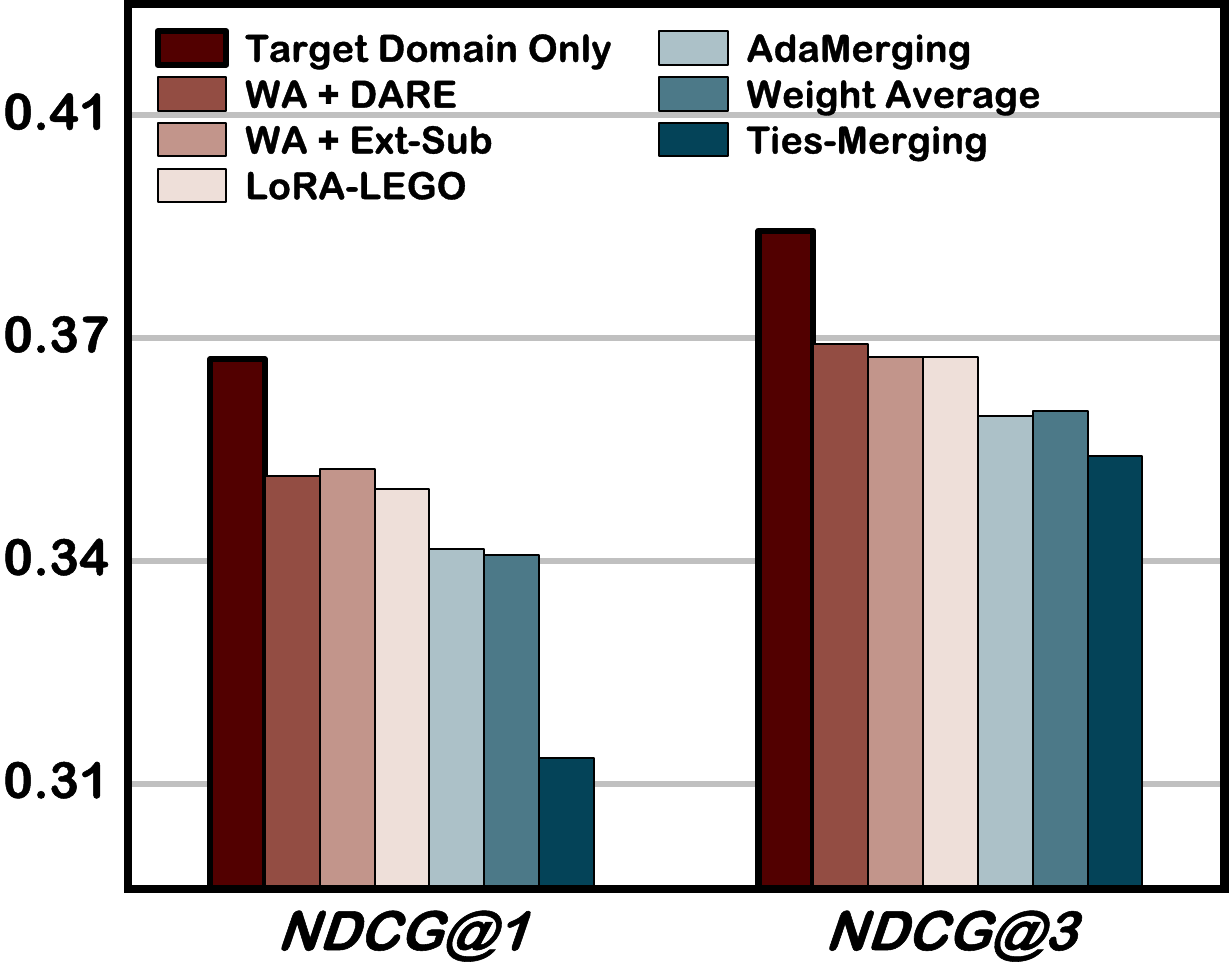}\label{fig:sub_a}}
    \hfil
    \subfloat[Different source domains]{\includegraphics[width=0.48\columnwidth]{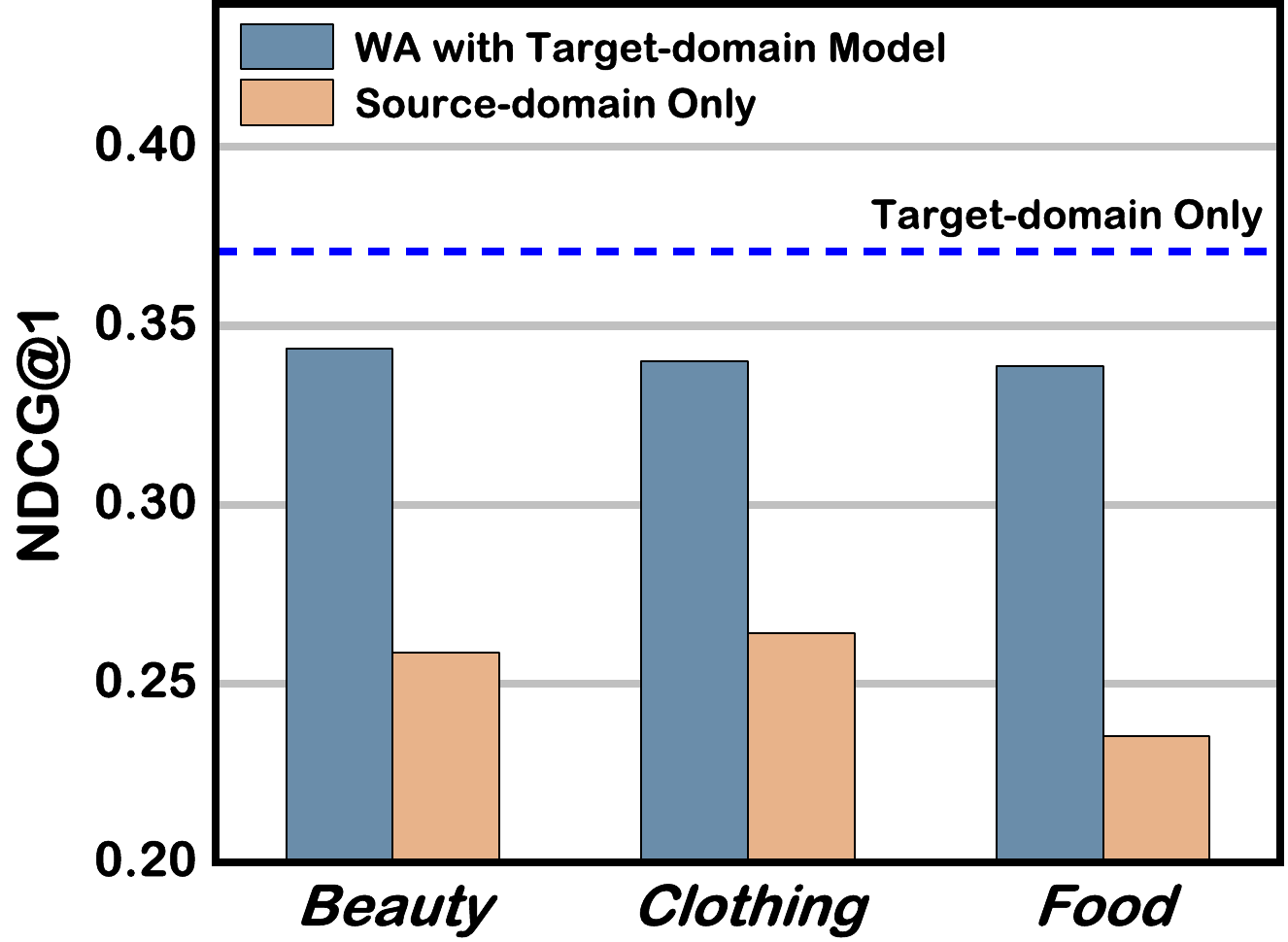}\label{fig:sub_b}}
    \caption{Performance comparison of different model merging methods when Sports is the target domain.} 
    %\vspace{-0.4cm}
    \label{fig:analysis} 
\end{figure}

\begin{figure}[ht]
    \centering
    \includegraphics[width=0.48\textwidth]{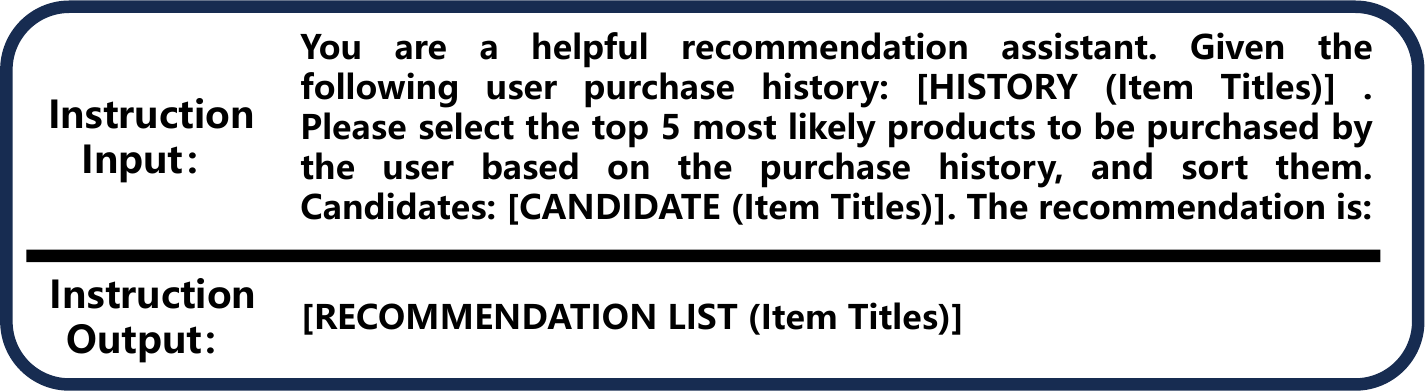}
    \vspace{-0.4cm}
    \caption{An example prompt of WeaveRec.}
    \vspace{-0.4cm}
    \label{fig:prompt}
    \Description{}
\end{figure}

\noindent$\bullet$ \quad\textbf{Analysis of the Experimental Observations.}
As shown in Figure \ref{fig:sub_a}, none of these methods can effectively enhance the target domain's knowledge. Their performance consistently falls short of the target-domain LoRA. This significant performance drop is likely due to a fundamental difference between the objectives of mainstream model merging methods and our task. Our aim is to enhance the performance of the merged model on the target domain by introducing models rich in recommendation knowledge from source domains, thereby reflecting the contribution of source domain knowledge to target domain improvement. Conversely, mainstream model merging methods are predominantly designed for multi-task scenarios. Their goal is to obtain a single model that achieves an acceptable performance across multiple tasks simultaneously. However, these performances are, in most cases, inferior to the performance of their respective single-task models.

Further experiments are shown in Figure \ref{fig:sub_b}. For the same target-~domain model, its fusion with different source domain models consistently leads to a significant decline in performance on the target domain. Concurrently, the performance of these source domain models on the target domain is notably poor, which is entirely expected, as source domain models have not been exposed to the target domain's training data. We can intuitively observe from Figure \ref{fig:analysis} that the performance of models obtained through various model merging methods lies between that of the target domain model and the source domain models. This implies that incorporating source-domain models degrades the overall performance to some extent, which aligns with findings from prior work~\cite{pmlr-v162-wortsman22a} suggesting that model merging should only include models exceeding a performance threshold.
Based on the findings above, our goal shifts to improving the performance of the source domain model on the target domain.
However, ensuring the source model's performance on the target domain is not trivial. The source and target domains often exhibit significant differences in user behavior patterns, item characteristics, and interaction distributions, making it difficult for source-domain models to generalize effectively to the target domain without substantial adaptation.
\iffalse
Consequently, we need to re-examine and propose novel merging method that allows the merged model to surpass the performance of the target-domain model on the target domain.
\fi

\section{Methodology}
\begin{figure}[!t]
    \centering
    \includegraphics[width=0.48\textwidth]{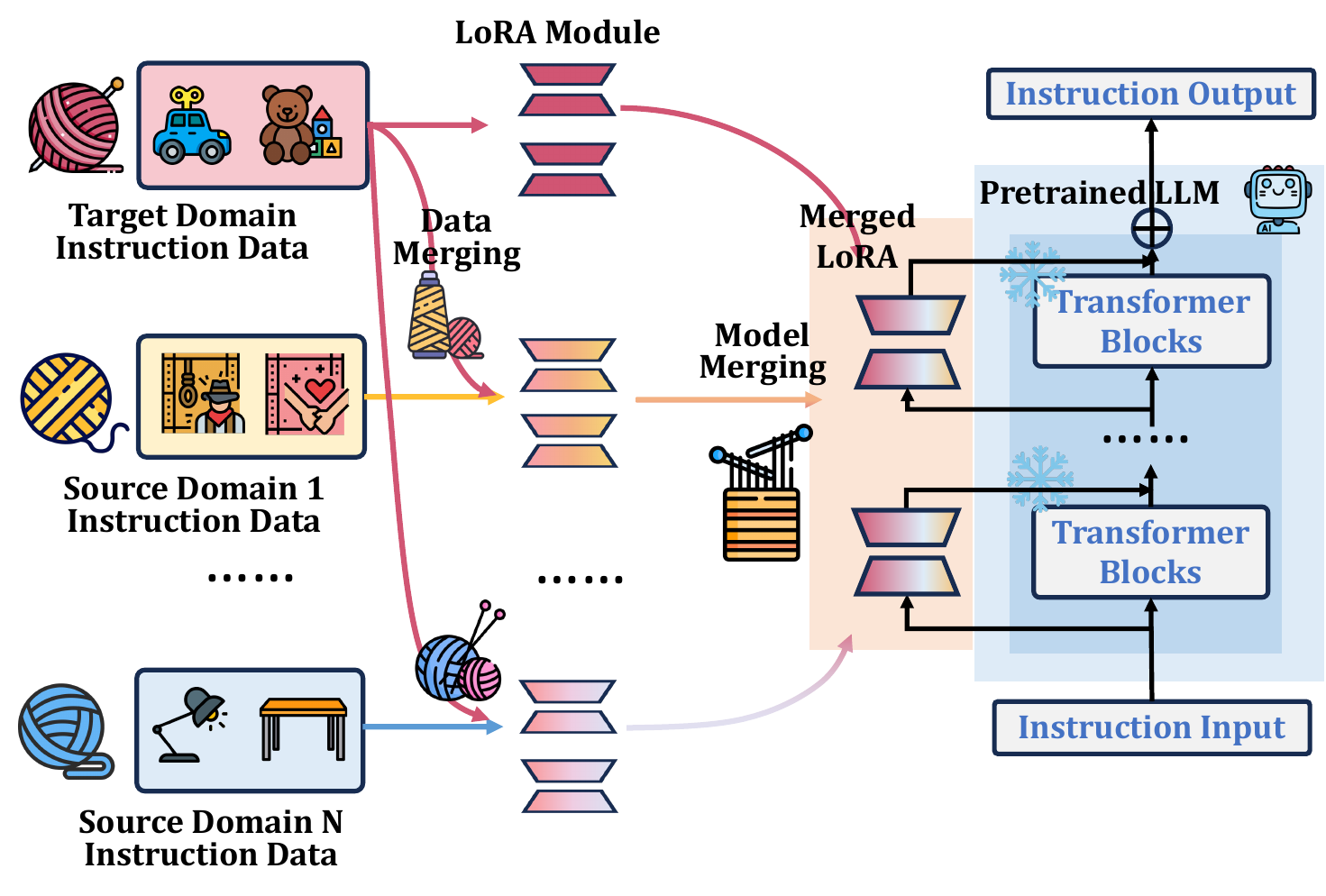}
    \vspace{-0.4cm}
    \caption{Illustration of our proposed WeaveRec Framework.}
    \vspace{-0.4cm}
    \label{fig:case}
    \Description{}
\end{figure}

In this section, we propose \textit{WeaveRec}, an effective and efficient framework of LoRA merging for LLM-based Cross-Domain Sequential Recommendation with mitigation of the performance degradation mentioned earlier. 

\subsection{WeaveRec} \label{msec1}
In this subsection, we introduce the proposed WeaveRec in detail.
As shown in figure \ref{fig:case}, WeaveRec comprises three stages. First, user data from all domains are processed and converted into instruction data to align the LLM with recommendation tasks. Second, for the initialized LoRA, we divide it into $N+1$ branches. The first branch is fine-tuned solely with target domain instruction data to obtain the \textbf{target-domain LoRA}. The remaining $N$ branches are fine-tuned by mixing target domain data with data from the $n$-th source domain $D_n$, respectively, to obtain \textbf{$N$ hybrid LoRAs}. In the final stage, we perform a model merging of the target-domain LoRA with the $N$ hybrid LoRAs. The merged LoRA is then loaded into the LLM, which is subsequently tested on the target domain. 

\noindent$\bullet$ \quad\textbf{Instruction Dataset Construction.} For a domain $D_n$ in the set of domains $\mathcal{D}=\{D_0, D_1, ..., D_N\}$, we design instruction templates to convert all user interaction sequences $s\in\mathcal{S}_n$ into textual instructions, as shown in Figure \ref{fig:prompt}. Notably our method doesn't demand intricate prompt engineering, highlighting its generality. Each instruction data $\mathcal{D}^I_n=\{(\textbf{x},\textbf{y})\}$ in the training dataset $\mathcal{X}_n$ of the domain $D_n$ contains the instruction input $\textbf{x}$ and output $\textbf{y}$. The instruction input includes a user's historical interactions, a set of item candidates, and the task description. The candidate set consists of one ground-truth item and some randomly selected negative samples. The instruction output is a ranked list of the user’s next most likely products to interact with. Note that all items within the instructions are represented by their titles to ensure transferability.

\noindent$\bullet$ \quad\textbf{Training Target-Domain Module.}To learn the specific knowledge in the target domain, we use the instruction dataset of the target domain $\mathcal{D}_0^I$ to train a LoRA module $\theta_0^*$:
\begin{equation}
    \theta_0^*=argmin_{\theta}\{-\sum_{(\mathbf{x,y})\in\mathcal{D}_0^I}\sum_{t=1}^{|\textbf{y}|}logP_{\Theta+\theta_0}(y^t\mid \textbf{y}^{<t}, \textbf{x})\}. \label{eq4tar-dom}
\end{equation}

\noindent$\bullet$ \quad\textbf{Training Hybrid Source-Domain Modules.} 
To extract domain-specific knowledge from individual source domains, we combine instruction data from the target domain with data from each source domain to train N corresponding source domain models. Specifically, for each source domain $n \in \{1, 2, \ldots, N\}$, we train a LoRA module $\theta_n^*$:
\begin{equation}
    \theta_n^*=argmin_{\theta_n}\{-\sum_{(\mathbf{x,y})\in\mathcal{D}_0^I \cup\mathcal{D}^I_n}\sum_{t=1}^{|\textbf{y}|}logP_{\Theta+\theta_n}(y^t\mid \textbf{y}^{<t}, \textbf{x})\}. \label{eq4}
\end{equation}

\noindent$\bullet$ \quad\textbf{Model Merging.} After instruction fine-tuning, a target-domain LoRA and $N$ hybrid source-domain LoRAs are obtained. For the single-task LoRA $\theta_0$ and all $N$ hybrid LoRAs $\{\theta_n\}_{n=1}^N$, we perform:
\begin{gather}
    \theta_\text{merged}=\sum_{i=0}^N\lambda_i\theta_i^*=\{\boldsymbol{B}_\text{merged}^l,\boldsymbol{A}_\text{merged}^l\}_{l=1}^L, \\
    \boldsymbol{B}_\text{merged}^l=\sum_{i=0}^N\lambda_i\boldsymbol{B}_i^l, \quad\boldsymbol{A}_\text{merged}^l=\sum_{i=0}^N\lambda_i\boldsymbol{A}_i^l,
\end{gather}
where the coefficients $\{\lambda_{i}\}_{i=0}^N$ represent the importance of corresponding branches and satisfy $\sum_{i=0}^N \lambda_i=1$. These coefficients can be treated as hyperparameters and determined through validation set tuning, or simply set to $\frac{1}{N+1}$.

Our method injects information from $N$ source domains into the target domain by leveraging model merging. It effectively mitigates the problem mentioned above and significantly enhances the model's performance on the target domain. Notably, Our framework is a plug-and-play solution. The training cost for these $N$ LoRAs is largely consistent, allowing for simultaneous or separate training, which demonstrates excellent scalability. Additionally, since our goal is multi-target cross-domain recommendation, which leverages data from multiple domains simultaneously to improve accuracy across all of them, each hybrid LoRA will be utilized twice, demonstrating high resource efficiency. For instance, considering the \textit{Sports-Clothing} hybrid LoRA, it will be utilized once when Sports is the target domain and Clothing is the source domain, and then again when the roles are reversed. This highlights WeavRec's ability to quickly adapt to source domain increase or decrease.

\subsection{Discussion} \label{motivation}
\noindent$\bullet$ \quad\textbf{Motivation.}
Based on the analysis from the Section \ref{analysis}, we hypothesize that the poor performance of the source domain model (as one of the merging components) on the target domain leads to the corruption of target domain knowledge. The merged model fails to effectively leverage source domain knowledge to enhance performance on the target domain; instead, it degrades the target domain model's original performance. \uline{Therefore, our intuitive idea is that all merging components should exhibit strong performance on the target domain to potentially yield enhancements. We aim to identify a model that can replace the source domain model for merging.} This model should possess two key characteristics: first, it should contain relatively rich recommendation knowledge from the source domain; and second, its performance on the target domain should be as strong as possible. \uline{Ultimately, we adopt a model trained with a mixture of source and target domain data to replace the source domain model for model merging.}

\begin{figure}[!t]
    \centering
    \includegraphics[width=1\columnwidth]{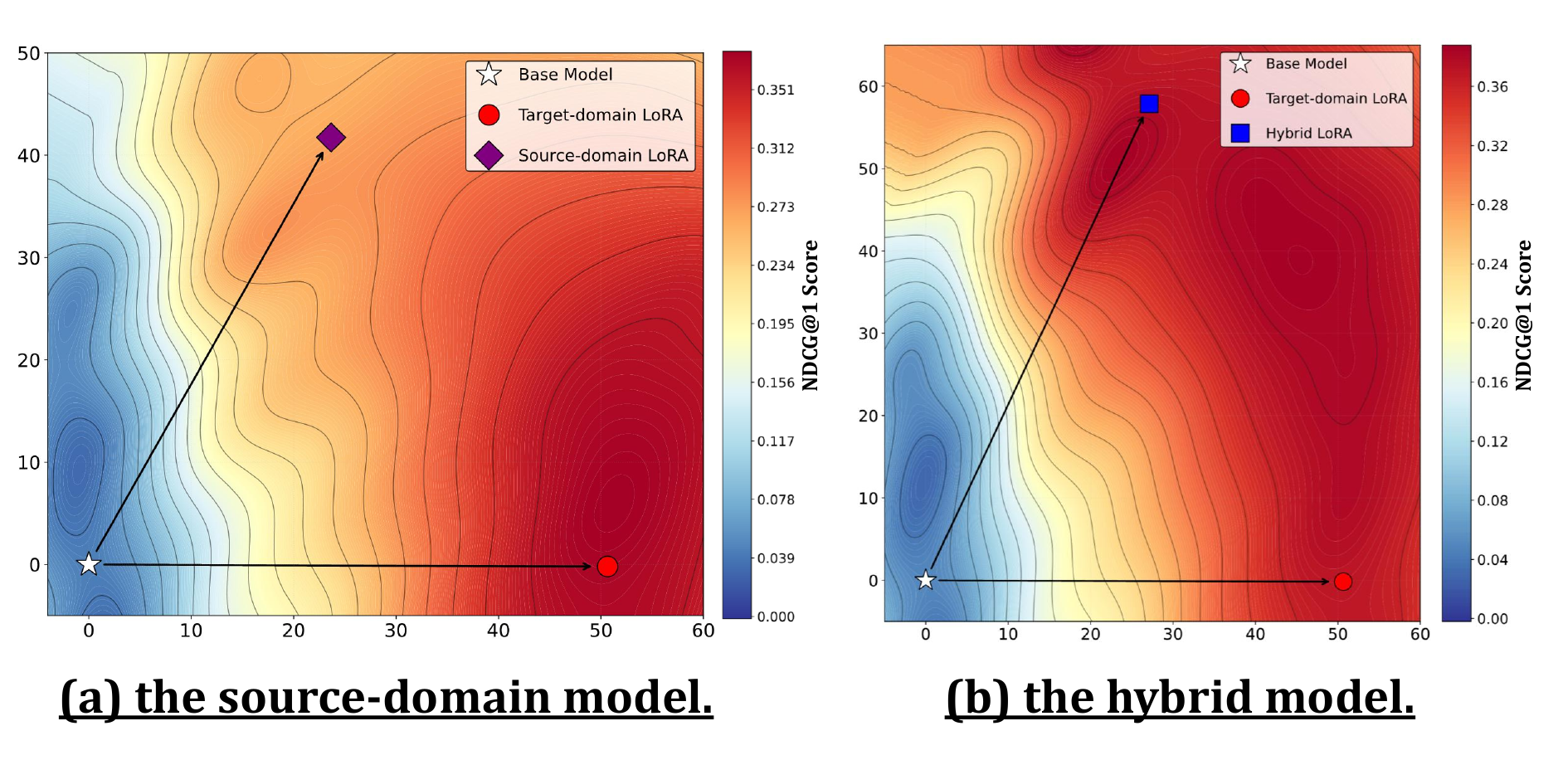}
    \caption{Landscape of test performance on the target domain. The target domain is Sports and the source domain is Clothing.} 
    \vspace{-0.4cm}
    \label{landscape} 
\end{figure}

\noindent$\bullet$ \quad\textbf{Loss Landscape Analysis.}
We further conducted a loss landscape analysis to verify our hypothesis.
In deep learning, the loss landscape describes how the loss changes with respect to different parameter configurations, and it reflects whether different models converge to similar or distant regions. When models lie in the same or nearby valleys of the landscape, their parameters can often be merged smoothly; otherwise, merging tends to hurt performance~\cite{pmlr-v162-wortsman22a,izmailov2018averaging}.
The performance landscape shown in Figure \ref{landscape}a indicates that the target-domain model is at the peak, while the source-domain model is at the foot of the mountain. The performance of the model merged from the two using existing methods is highly likely to fall in the region between them, and it is difficult to push it to a higher peak. As shown in Figure \ref{landscape}b, when the source-domain model is replaced by the hybrid model from WeaveRec, both are at the peak. Merging them, the performance of the merged model tends to reach a higher region. Therefore, Figure \ref{landscape} echoes that the members involved in the fusion should achieve relatively high accuracy on the target domain.

\noindent$\bullet$ \quad\textbf{Theoretical Analysis.} Furthermore, from the perspective of domain adaptation theory, we can analyze that the generalization error upper bound of this dual-domain mixed-training model is lower than that of the source domain model. Let $D_T$, $D_S$, and $D_M$ denote the target, source, and mixed distributions, respectively, where:
\begin{equation}
    D_M\sim (1+Z)D_T+ZD_S,\ Z\sim Bernoulli(\frac{\lambda}{1+\lambda}),\ \lambda\neq 0.
\end{equation}
Note that $\lambda=1$ indicates an equal mixing ratio of data from the two domains. We denote $h_S$ as the optimal hypothesis on distribution $D_S$, and $h_M$ as that on distribution $D_M$.
According to \cite{10.1007/s10994-009-5152-4}, the generalization error of the hypothesis $h_S$ on distribution $D_T$ is:
\begin{equation}
    \epsilon_T(h_S)=\mathbb{E}_{\textbf{x}\sim D_T}[|h_S(\textbf{x})-f^*_T(\textbf{x})|].
\end{equation}
where $f^*_T(\cdot)$ is the ground-truth function of distribution $D_T$. Likewise, we have:
\begin{equation}
    \epsilon_T(h_M)=\mathbb{E}_{\textbf{x}\sim D_T}\left[\left|h_M(\textbf{x})-f^*_T(\textbf{x})\right|\right].
\end{equation}
We can represent the upper bounds of $\epsilon_T(h_S)$ and $\epsilon_T(h_M)$ with the following two inequalities \cite{10.1007/s10994-009-5152-4}, respectively:
\begin{gather}
    \epsilon_T(h_S)\leq \epsilon_S(h_S)+d_{\mathcal{H}}(D_S, D_T)+\lambda^* \label{ineq_1}, \\
    \epsilon_T(h_M)\leq \epsilon_S(h_M)+d_{\mathcal{H}}(D_M, D_T)+\lambda^* \label{ineq_2},
\end{gather}
where $\lambda^*$ refers to a constant related to ground-truth functions and $d_{\mathcal{H}}$ is a concept known as H-divergence. The definition of ${d_\mathcal{H}}$ is:
$$
d_{\mathcal{H}}(D_1,D_2) = 2 \sup_{h \in \mathcal{H}} \left| P_{\textbf{x} \sim D_1}\left[h(x)=1\right] - P_{\textbf{x} \sim D_2}\left[h(x)=1\right] \right|.
$$
This supremum formula characterizes the distance between distributions $D_1$ and $D_2$ by finding the best function $h$ in the function space $\mathcal{H}$ such that the probability of successful prediction on distribution $D_1$ is maximized, and the probability of successful prediction on distribution $D_2$ is minimized. 
Since $D_M$ has inherent overlap with $D_T$, any optimal function distinguishing $D_M$ from $D_T$ must correctly predict samples from both distributions, leading to $d_{\mathcal{H}}(D_M, D_T) < d_{\mathcal{H}}(D_S, D_T)$. Given that $\epsilon_S(h_S) \approx \epsilon_M(h_M)$ for converged models, we conclude:
\begin{equation}
    Bound(\epsilon_T(h_M)) < Bound(\epsilon_T(h_S)).
\end{equation}
The model $h_M$ possesses a lower generalization error upper bound on $D_T$. This indicates its error on the target domain is more controllable, leading to relatively better performance compared to source domain model $h_S$. 

\noindent$\bullet$ \quad\textbf{Efficiency Analysis.}
After merging all LoRA modules, we retain only a single LoRA module. As a result, there is no additional memory or computational overhead during inference.
WeaveRec offers plug-and-play integration, where a newly arriving source domain can be seamlessly accommodated by simply training one additional hybrid LoRA module, without the need to retrain or modify the existing ones. This design ensures both scalability and efficiency when adapting to diverse domains.

\iffalse
This naturally leads into the next section, where we present our formal methodology.
\fi

\section{Experiments}
\begin{table*}[htbp] % table* 用于创建跨双栏的表格
    \centering
    \caption{Performance comparison in cross-domain scenario.}
    \label{tab:cross-domain}
    \resizebox{\textwidth}{!}{ % \resizebox 可以强制表格适应文本宽度
    \setcellgapes{3pt} % 调整单元格内部的垂直间距，可根据需要调整
    \makegapedcells % 启用 \setcellgapes
    \begin{tabular}{l|cccc|cccc|cccc|cccc} % 列格式：l左对齐，c居中对齐，|绘制垂直线
        \toprule % 表格顶部粗线
        \multicolumn{1}{l|}{\multirow{2}{*}{\textbf{Method}}} & \multicolumn{4}{c|}{\textbf{Beauty,Clothing,Food$\to$Sports}} & \multicolumn{4}{c|}{\textbf{Sports,Clothing,Food$\to$Beauty}} & \multicolumn{4}{c|}{\textbf{Beauty,Sports,Food$\to$Clothing}} & \multicolumn{4}{c}{\textbf{Beauty,Sports,Clothing$\to$Food}} \\
        \cmidrule(lr){2-5} \cmidrule(lr){6-9} \cmidrule(lr){10-13} \cmidrule(lr){14-17} % 绘制部分横线
        & N@1 & N@3 & N@5 & M@5 & N@1 & N@3 & N@5 & M@5 & N@1 & N@3 & N@5 & M@5 & N@1 & N@3 & N@5 & M@5 \\
        \midrule % 表格中部细线
        \textbf{GRU4Rec} & 0.1664 & 0.2635 & 0.3139 & 0.2776 & 0.1737 & 0.2748 & 0.3378 & 0.2901 & 0.1531 & 0.2471 & 0.2894 & 0.2466 & 0.1823 & 0.2763 & 0.3554 & 0.3041 \\
        \textbf{SASRec} & 0.1890 & 0.3142 & 0.3699 & 0.3132 & 0.2166 & 0.3272 & 0.3712 & 0.3239 & 0.1963 & 0.3193 & 0.3714 & 0.3069 & 0.2366 & 0.3533 & 0.3979 & 0.3492 \\
        \textbf{BERT4Rec} & 0.1839 & 0.2792 & 0.3205 & 0.2779 & 0.2215 & 0.3116 & 0.3447 & 0.3073 & 0.0813 & 0.1415 & 0.1748 & 0.1444 & 0.2440 & 0.3405 & 0.3724 & 0.3343 \\
        \textbf{FMLP-Rec} & 0.2411 & 0.3462 & 0.3879 & 0.3432 & 0.2581 & 0.3613 & 0.4003 & 0.3598 & 0.1842 & 0.2691 & 0.3084 & 0.3136 & 0.2934 & 0.3873 & 0.4197 & 0.4084 \\
        \midrule % 表格中部细线
        \textbf{MCRPL} & 0.2465 & 0.3330 & 0.3775 & 0.3523 & 0.2424 & 0.3490 & 0.3912 &  0.3677 & 0.2034 & 0.2967 & 0.3527 & 0.3162 & 0.2473 & 0.3577 & 0.4009 & 0.3539 \\
        \textbf{UnisRec} & 0.2258 & 0.3323 & 0.3764 & 0.3299 & 0.2485 & 0.3367 & 0.3727 & 0.3345 & 0.1948 & 0.2936 & 0.3371 & 0.2927 & 0.2965 & 0.3808 & 0.4122 & 0.3773 \\
        \textbf{VQ-Rec} & 0.2512 & 0.3578 & 0.3812 & 0.3550 & 0.2686 & 0.3674 & 0.3924 & 0.3498 & 0.2367 & 0.3562 & 0.3895 & 0.3327 & 0.3104 & 0.3828 & 0.4131 & 0.3945 \\
        \textbf{RecFormer} & 0.2638 & 0.3575 & 0.3816 & 0.3694 & 0.2844 & 0.3751 & 0.4160 & 0.3831 & 0.2568 & 0.3586 & 0.3792 & 0.3451 & 0.3125 & 0.3918 & 0.4376 & 0.3956 \\
        \midrule
        \textbf{Qwen2-7B} & 0.0411 & 0.0488 & 0.0659 & 0.0560 & 0.0450 & 0.0559 & 0.0728 & 0.0623 & 0.0730 & 0.0880 & 0.1087 & 0.0955 & 0.0282 & 0.0366 & 0.0516 & 0.0426 \\
        \textbf{TALLRec} & 0.2957 & 0.3232 & 0.3435 & 0.3272 & 0.2604 & 0.2885 & 0.3078 & 0.3570 & 0.3124 & 0.3403 & 0.3593 & 0.3434 & 0.3184 & 0.3445 & 0.3627 & 0.3477 \\
        \textbf{LLM-REC} & 0.3206 & 0.3896 & \underline{0.4107} & \underline{0.4059} & 0.3623 & \underline{0.4305} & \underline{0.4478} & \underline{0.4329} & 0.3227 & 0.3812 & \underline{0.4076} & 0.3854 & 0.3475 & 0.4078 & 0.4522 & \underline{0.4217} \\
        \midrule
        \textbf{Weight Average} & 0.3098 & 0.3321 & 0.3510 & 0.3368 & 0.3369 & 0.3595 & 0.3766 & 0.3629 & 0.3269 & 0.3512 & 0.3689 & 0.3544 & 0.3028 & 0.3275 & 0.3460 & 0.3311 \\
        \textbf{AdaMerging} & 0.3095 & 0.3326 & 0.3502 & 0.3361 & 0.3384 & 0.3609 & 0.3782 & 0.3645 & 0.3270 & 0.3512 & 0.3689 & 0.3545 & 0.3025 & 0.3275 & 0.3468 & 0.3315 \\
        \textbf{LoRA-LEGO} & 0.3109 & 0.3328 & 0.3502 & 0.3365 & 0.3481 & 0.3704 & 0.3867 & 0.3733 & 0.3244 & 0.3482 & 0.3655 & 0.3514 & 0.3084 & 0.3320 & 0.3493 & 0.3354 \\
        \textbf{Ties-Merging} & 0.3102 & 0.3361 & 0.3555 & 0.3400 & 0.3375 & 0.3402 & 0.3420 & 0.3405 & 0.3275 & 0.3543 & 0.3747 & 0.3585 & 0.2994 & 0.3250 & 0.3444 & 0.3289 \\
        \midrule
        \textbf{Target-domain Only} & \underline{0.3708} & 0.3904 & 0.4057 & 0.3936 & \underline{0.4071} & 0.4293 & 0.4438 & 0.4314 & \underline{0.3643} & \underline{0.3880} & 0.4049 & \underline{0.3910} & 0.4143 & 0.4337 & 0.4492 & 0.4370 \\
        \textbf{All Data Merging} & 0.3677 & \underline{0.3919} & 0.4092 & 0.3950 & 0.2965 & 0.3231 & 0.3413 & 0.3260 & 0.3545 & 0.3783 & 0.3963 & 0.3818 & \underline{0.4146} & \underline{0.4375} & \underline{0.4554} & 0.4412 \\
        \textbf{WeaveRec (ours)} & \textbf{0.3897*} & \textbf{0.4107*} & \textbf{0.4253*} & \textbf{0.4132*} & \textbf{0.4180*} & \textbf{0.4386*} & \textbf{0.4543*} & \textbf{0.4418*} & \textbf{0.3732*} & \textbf{0.3965*} & \textbf{0.4130*} & \textbf{0.3995*} & \textbf{0.4220*} & \textbf{0.4425*} & \textbf{0.4572*} & \textbf{0.4452*} \\
        \bottomrule
    \end{tabular}
    } % 结束 \resizebox
\end{table*}

\begin{table*}[htbp] % table* 用于创建跨双栏的表格
    \centering
    \caption{Performance comparison in cross-platform scenario.}
    \label{tab:cross-platform}
    \resizebox{0.7\textwidth}{!}{ % \resizebox 可以强制表格适应文本宽度
    \setcellgapes{3pt} % 调整单元格内部的垂直间距，可根据需要调整
    \makegapedcells % 启用 \setcellgapes
    \begin{tabular}{l|cccc|cccc} % 列格式：l左对齐，c居中对齐，|绘制垂直线
        \toprule % 表格顶部粗线
        \multicolumn{1}{c|}{\multirow{2}{*}{\textbf{Method}}} & \multicolumn{4}{c|}{\textbf{Toys$\to$MovieLens-1M}} & \multicolumn{4}{c}{\textbf{MovieLens-1M$\to$Toys}} \\
        \cmidrule(lr){2-5} \cmidrule(lr){6-9} % 绘制部分横线
        & NDCG@1 & NDCG@3 & NDCG@5 & MRR@5 & NDCG@1 & NDCG@3 & NDCG@5 & MRR@5 \\
        \midrule % 表格中部细线
        \textbf{GRU4Rec} & 0.2211 & 0.3750 & 0.4419 & 0.3729 & 0.1548 & 0.2524 & 0.2987 & 0.2531 \\
        \textbf{SASRec} & 0.2754 & 0.3743 & 0.4339 & 0.3662 & 0.2081 & 0.3157 & 0.3591 & 0.3127 \\
        \textbf{BERT4Rec} & 0.2405 & 0.3682 & 0.4256 & 0.3678 & 0.1508 & 0.2334 & 0.2732 & 0.2345 \\
        \textbf{FMLP-Rec} & 0.2853 & 0.4378 & 0.4788 & 0.4458 & 0.2614 & 0.3562 & 0.3919 & 0.3707\\
        \midrule % 表格中部细线
        \textbf{MCRPL} & 0.2911 & 0.3807 & 0.4323 & 0.4016 & 0.2378 & 0.3572 & 0.3889 & 0.3551 \\
        \textbf{UnisRec} & 0.3011 & 0.4325 & 0.4810 & 0.4266 & 0.2318 & 0.3373 & 0.3792 & 0.3340 \\
        \textbf{VQ-Rec} & 0.3362 & 0.4569 & 0.4945 & 0.4334 & 0.2641 & 0.3666 & 0.3982 & 0.3616 \\
        \textbf{RecFormer} & 0.2847 & 0.4309 & 0.4795 & 0.4252 & 0.3012 & 0.3872 & 0.4188 & 0.3804 \\
        \midrule
        \textbf{Qwen2-7B} & 0.0099 & 0.0135 & 0.0145 & 0.0132 & 0.0955 & 0.1146 & 0.1326 & 0.1197\\
        \textbf{TALLRec} & 0.2972 & 0.3177 & 0.3331 & 0.3208 & 0.3174 & 0.3456 & 0.3661 & 0.3496 \\
        \textbf{LLM-REC} & 0.4023 & 0.4766 & \underline{0.4952} & 0.4701 & 0.3238 & 0.4209 & 0.4452 & 0.4255 \\
        \midrule
        \textbf{Weight Average} & 0.4103 & 0.4327 & 0.4505 & 0.4367 & 0.3595 & 0.3843 & 0.4021 & 0.3875\\
        \textbf{AdaMerging} & 0.4111 & 0.4334 & 0.4515 & 0.4375 & 0.3596 & 0.3849 & 0.4024 & 0.3879\\
        \textbf{LoRA-LEGO} & 0.4081 & 0.4291 & 0.4442 & 0.4319 & 0.3704 & 0.3887 & 0.4035 & 0.3920\\
        \textbf{Ties-Merging} & 0.1930 & 0.2637 & 0.2951 & 0.2633 & 0.2942 & 0.3475 & 0.3723 & 0.3479\\
        \midrule
        \textbf{Target-domain Only} & 0.4500 & 0.4704 & 0.4845 & 0.4728 & \underline{0.4080} & \underline{0.4328} & \underline{0.4488} & \underline{0.4350} \\
        \textbf{All Data Merging} & \underline{0.4568} & \underline{0.4752} & 0.4888 & \underline{0.4777} & 0.3984 & 0.4239 & 0.4423 & 0.4274\\
        \textbf{WeaveRec (ours)} & \textbf{0.4854*} & \textbf{0.5049*} & \textbf{0.5217*} & \textbf{0.5073*} & \textbf{0.4110*} & \textbf{0.4368*} & \textbf{0.4541*} & \textbf{0.4396*} \\
        \bottomrule
    \end{tabular}
    } % 结束 \resizebox
\end{table*}

\subsection{Experimental Settings}
\subsubsection{\textbf{Datasets.}} We conduct experiments on two scenarios to demonstrate the generalization capability of our method: \textbf{cross-domain} scenario and \textbf{cross-platform} scenario. 

For the cross-domain scenario, we select four e-commerce domains in Amazon (Beauty, Sports, Clothing, and Food). Duo to our goal is multi-target CDSR, we denote $Beauty,Clothing,Food\to Sports$ to signify that $Sports$ is the target domain, while $Beauty$ $Clothing$ $Food$ are source domains. This arrangement leads to four different experimental setups, designed to leverage data from all four domains to enhance the model's performance across each of them. For the cross-platform scenario, we select the Amazon Toys and MovieLens-1M, originating from distinct platforms. Similarly, we have two types of experimental setups: $MovieLens1M\to Toys$ and $Toys\to MovieLens1M$.

For all datasets, items are represented using their textual "title" information. We keeps the five-core data and filters out users and items with fewer than five interactions for all datasets. Following \cite{p5, 2024bridging}, we adopt the leave-one-out strategy to split the filtered datasets, which split the last interaction of each user into the test set, the second-to-last one into the validation set, and the rest into the training set. Details of datasets can be found in Appendix \ref{appendix A.3}.

\subsubsection{\textbf{Baselines.}} To validate the effectiveness of WeaveRec, we compare it with five groups of baselines.
\textbf{1) Single-Domain Sequential Recommendation:} GRU4Rec~\cite{gru4rec}, SASRec~\cite{sasrec}, BERT4Rec \cite{bert4rec}, and FMLP-Rec~\cite{fmlp}.
\textbf{2) Cross-Domain Sequential Recommendation:} 
MCRPL~\cite{mcrpl}, VQ-Rec~\cite{hou2023vqrec}, UniSRec~\cite{unisrec}, and RecFormer~\cite{recformer}.
\textbf{3) LLM-Based Recommendation:} Qwen2-7B\footnote{\url{https://huggingface.co/Qwen/Qwen2-7B-Instruct}}, TALLRec~\cite{tallrec}, and LLM-Rec~\cite{one-model-for-all}.
\textbf{4) Model Merging Methods:}
Weight Average, AdaMerging~\cite{AdaMerging_ICLR_2024}, LoRA-LEGO \cite{lego}, and Ties-Merging~\cite{ties}.
\textbf{5) Our Ablation Counterparts:}
Target-Domain Only and All Data Merging. See Appendix \ref{appendix A.4} for more details of these baselines.

\subsubsection{\textbf{Evaluation Setting.}}\label{eval_part} Following some previous LLM-based recommendation works, to evaluate the performance of each methods, each user's candidate set in the test set includes 29 randomly selected non-interacted items and one ground truth item. To quantitatively compare, we employ widely used ranking-based metrics, NDCG@1, NDCG@3, NDCG@5, and MRR@5 for all experiments. All metrics show improved performance with higher values. For all the following tables, \textbf{bold*} numbers refer to the best performance, while \underline{underlined} numbers indicate the second-best performance.

 \subsection{Overall Performance}
The experimental results in the cross-domain scenario and cross-platform scenario are shown in Table \ref{tab:cross-domain} and \ref{tab:cross-platform}, respectively. The proposed \textit{WeaveRec} consistently achieves the best performance across various target domain settings in both cross-domain and cross-platform scenarios, with a t-test at p<0.05 level. From the experimental results, we have two main observations:

\noindent $\bullet$ Across both cross-domain and cross-platform scenarios, existing model merging methods underperform compared to target-domain-only models, whereas WeaveRec successfully leverages source-domain knowledge to achieve enhanced performance in the target domain. Notably, the four baseline model merging approaches demonstrate comparable performance across various experimental settings, with the exception of Ties-Merging which exhibits significant performance degradation in cross-platform scenarios. This finding suggests that current model merging methodologies are not properly aligned with the requirements of cross-domain recommendation tasks.

\noindent $\bullet$ Comparing the two baselines of target-domain only and all data merging, we can find that when using the merging of multi-domain data to train the model, it may performs better in the Sports and Food domains than the target-domain-only model, but its performance declines in Beauty and Clothing, especially in Beauty. This indicates the instability of multi-task joint training in cross-domain recommendation. When the model learns multi-domain data, the updated gradients have conflicting directions, which compromises the performance of the model. This may be manifested as the model performing better than the target-domain-only model in some domains, but experiencing severe degradation in some domains.

\subsection{In-Depth Analysis}
More in-depth analysis can be found in Appendix \ref{more-analysis}.

\subsubsection{\textbf{Component Analysis.}}
We disassemble \textit{WeaveRec} under the settings of Table \ref{tab:cross-domain} and individually test each component on target domains. Note that different target domains utilize different hybrid LoRAs, which are not distinguished in this section. As shown in Table \ref{tab:component-analysis}, the performance of each individual branch of \textit{WeaveRec} is comparable, while the performance of the merged model is significantly enhanced. This clearly demonstrates \textit{WeaveRec}'s mitigation of performance degradation. 
This supports our discussion in Section \ref{motivation} that fusion members should not exhibit poor performance on the target domain.

\subsubsection{\textbf{Why WeaveRec Employs Weight Average?}} An interesting question arises: Why does WeaveRec employ simple Weight Averaging (WA) for model merging instead of more complex methods like LoRA-LEGO or Tie-Merging? To explore this, we designate Sports as the target domain, with Clothing, Beauty, and Food serving as source domains. Following Our WeaveRec, we obtain three hybrid LoRAs: \textit{Sports-Clothing}, \textit{Sports-Beauty}, and \textit{Sports-Food}. We then merge these three LoRAs with the target-domain LoRA using Weight Average, LoRA-LEGO, Tie-Merging and AdaMerging as the merging functions. The results, as shown in Table \ref{tab:why-WA}, indicate that LoRA-LEGO and Tie-Merging still perform poorly, with Weight Average showing the best performance and AdaMerging being the second-best. This further suggests that existing model merging methods may not be well-suited for recommendation tasks. It's possible these methods manipulate model parameters too aggressively, leading to a loss of valuable knowledge.

\subsubsection{\textbf{Impact of the Number of Source Domains.}} As \textit{WeaveRec} supports plug-and-play modules, we can select one or more source domains to facilitate cross-domain recommendations for a target domain. To explore how the varying number of source domains affects the LLM's performance on target domain, we conduct four sets of experiments on two distinct target domains. \textit{WeaveRec-0} represents the baseline, degenerating to the target-domain LoRA, while \textit{WeaveRec-N} ($1\leq N\leq3$) signifies utilizing $N$ source domains for cross-domain recommendation.

\begin{figure}[!t]
    \centering
    \subfloat[Target Domain: Sports]{\includegraphics[width=0.49\columnwidth]{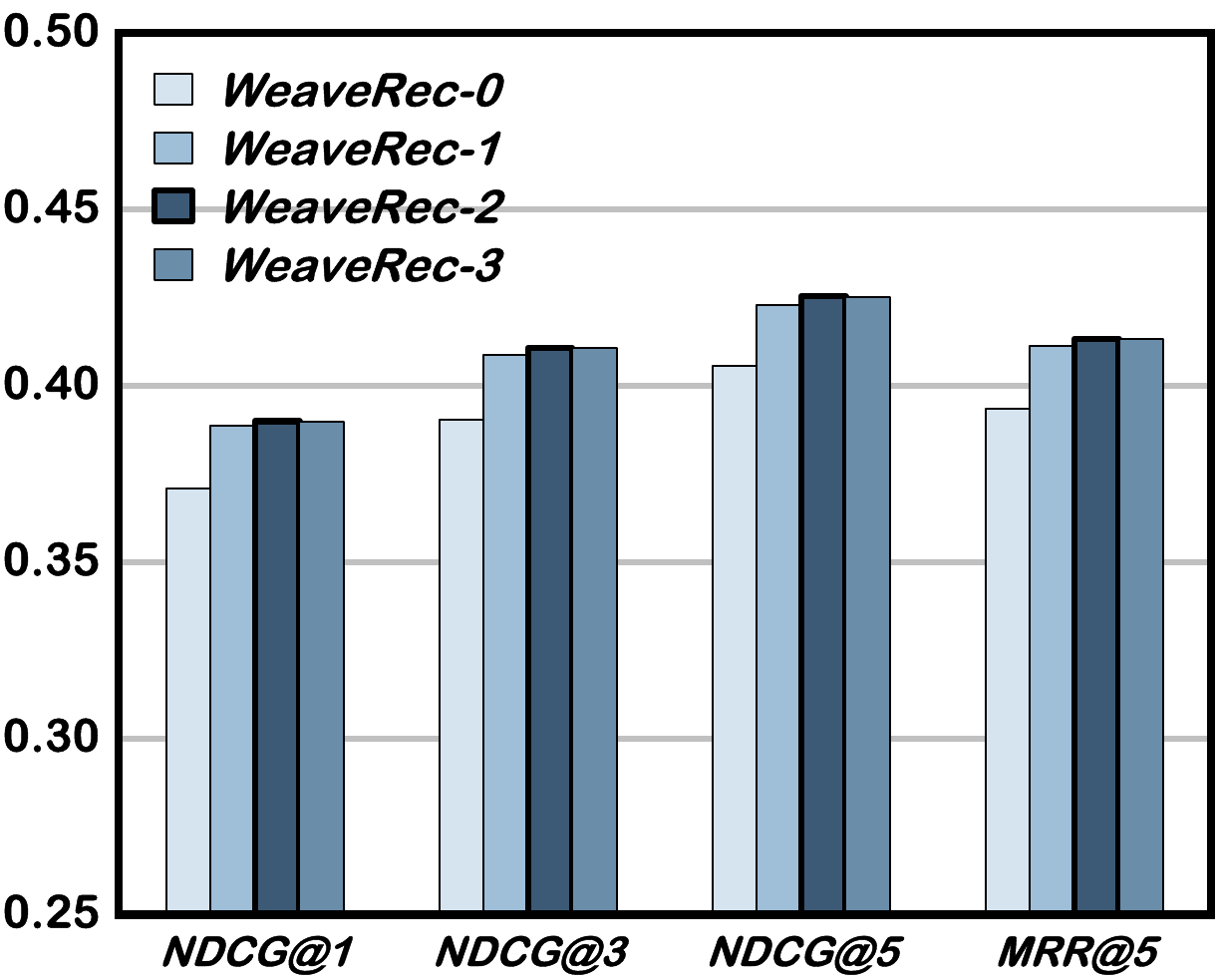} \label{indepth-3-a}}
    \hfil
    \subfloat[Target Domain: Beauty]{\includegraphics[width=0.49\columnwidth]{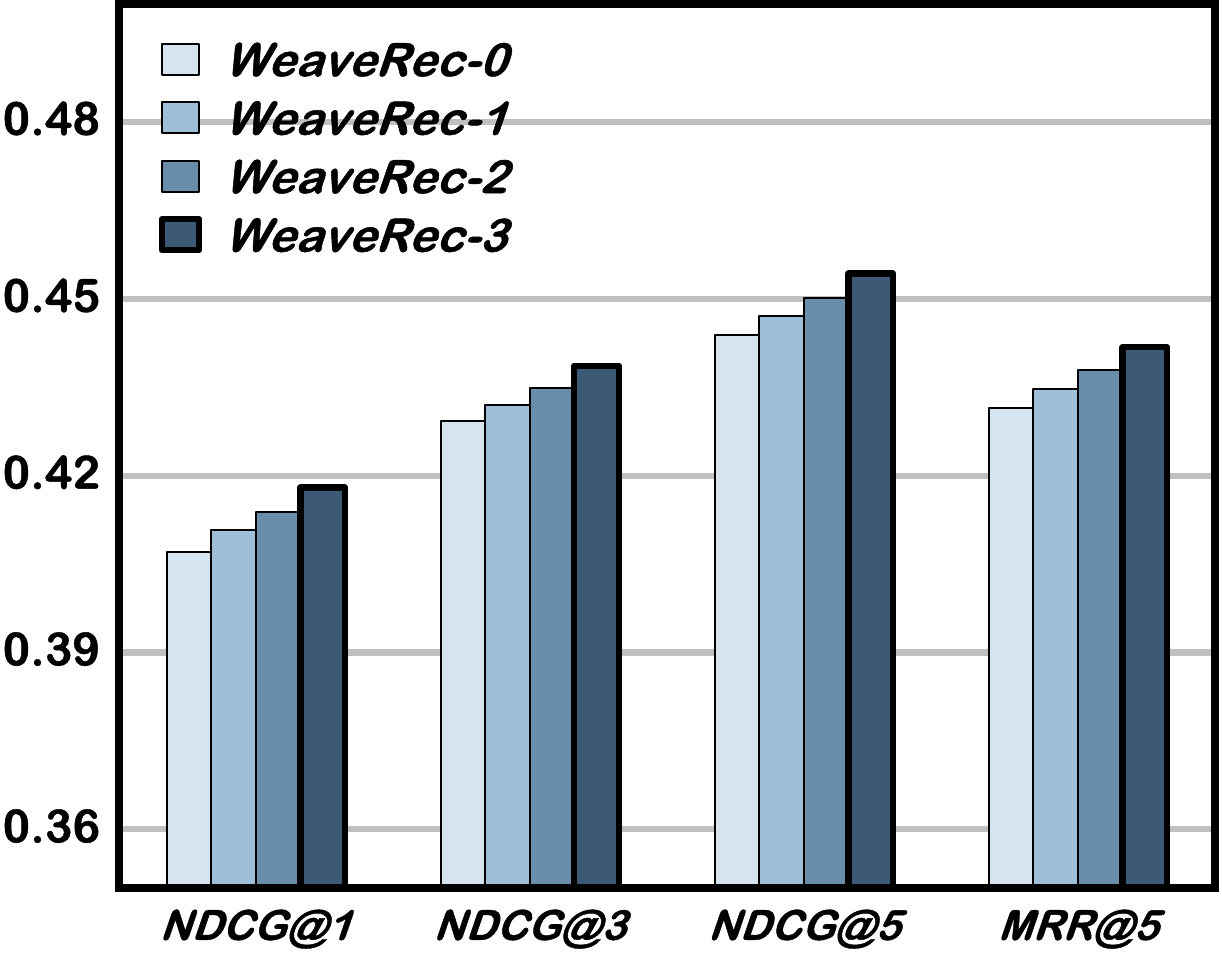} \label{indepth-3-b}}
    \vspace{-0.2cm}
    \caption{The impact of the number of source domains on \textit{WeaveRec}'s performance. (a) The three source domains are Amazon Clothing, Beauty, and Food. \textit{WeaveRec-N} refers to using only the first \textit{N} of these three domains. (b) Similarly, three source domains are Amazon Food, Clothing, and Sports.}
    \label{indepth-3} % 主图标签
\end{figure}

As shown in Figure~\ref{indepth-3-a}, for Sports as the target domain, the performance of \textit{WeaveRec-0}, \textit{WeaveRec-1}, and \textit{WeaveRec-2} exhibits little difference. This indicates that for the Sports domain, more source domains are not necessarily better. One source domain can effectively enhance the model's performance on Sports, with the addition of more source domains not yielding a substantial increase. Notably, even though \textit{WeaveRec-2} shows the best performance, adding another source domain does not lead to a significant performance drop. This suggests that while multiple source domains might not provide optimal enhancement for a particular target domain, they do not cause significant performance degradation, demonstrating \textit{WeaveRec}'s robustness.

In contrast, for Beauty as target domain, as presented in Figure~\ref{indepth-3-b}, we can intuitively observe a gradual increase in the model's test performance on the target domain with the increasing number of source domains. This implies that for the Beauty domain, utilizing three source domains yields the best results. This pattern differs from that observed in Figure \ref{indepth-3-a}, signifying that the optimal number of source domains varies across different target domains, thereby reflecting the inherent heterogeneity among them.

\iffalse
\subsubsection{\textbf{Impact of Sampling Ratio}} 
Here, we investigate the impact of the sampling ratio for fine-tuning hybrid LoRA on model performance. We designate Sports as the target domain and Clothing as one source domain. For fine-tuning the hybrid LoRA, we sampled $\alpha\%$ data from each of these two domains, where $\alpha\in\{10,25,40,55\}$. As illustrated in Figure \ref{indepth-2}, \textit{WeaveRec}'s performance already surpasses that of Target-domain LoRA when the sampling ratio exceeds approximately 55\%, demonstrating \textit{WeaveRec}'s high efficiency.
\fi

\subsubsection{\textbf{Sensitivity Analysis of Interpolation Weight.}} 
\begin{figure}[!t]
    \centering
    \subfloat[Sports $\to$ Clothing]{\includegraphics[width=0.48\columnwidth]{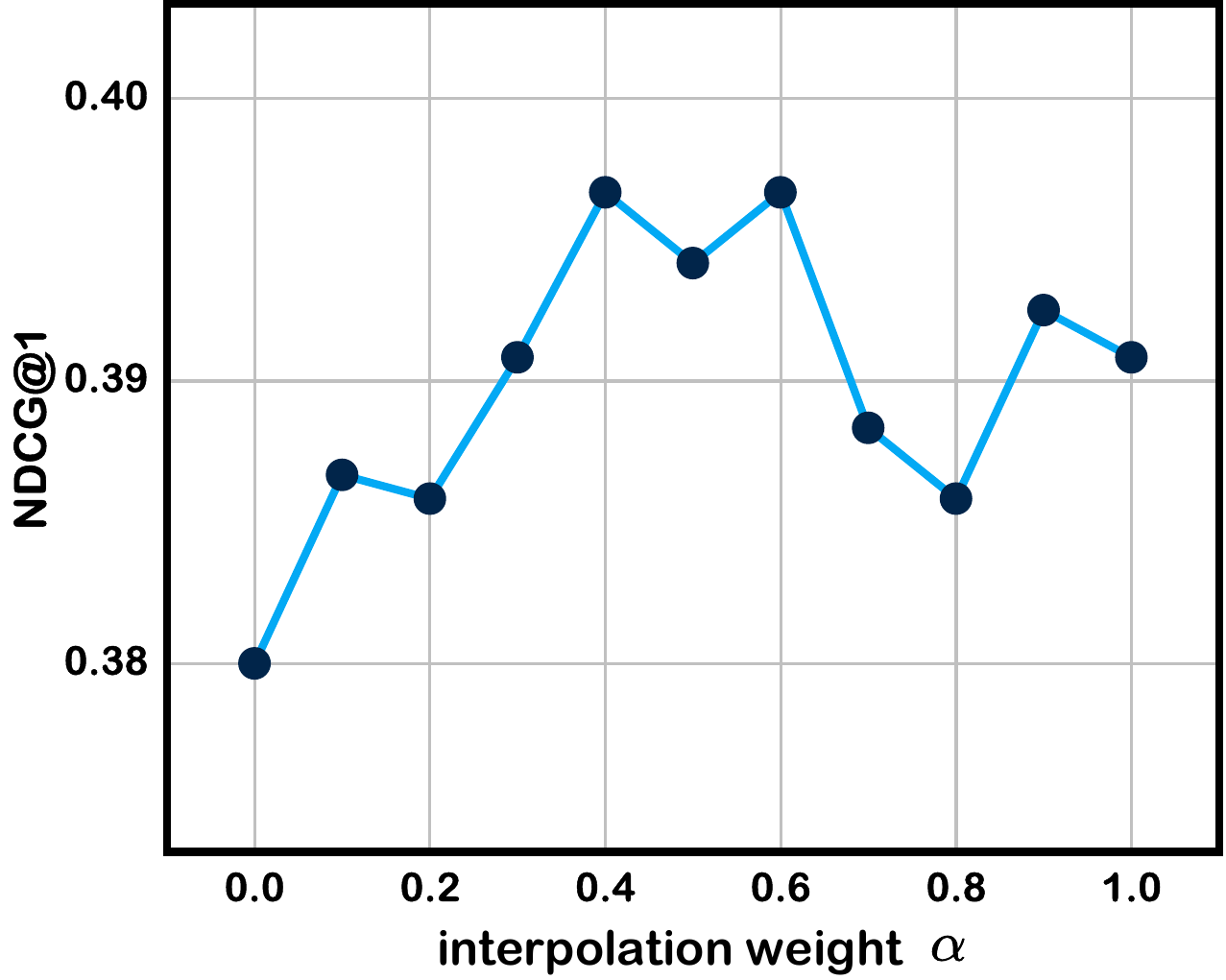}}
    \hfil
    \subfloat[Food $\to$ Beauty]{\includegraphics[width=0.49\columnwidth]{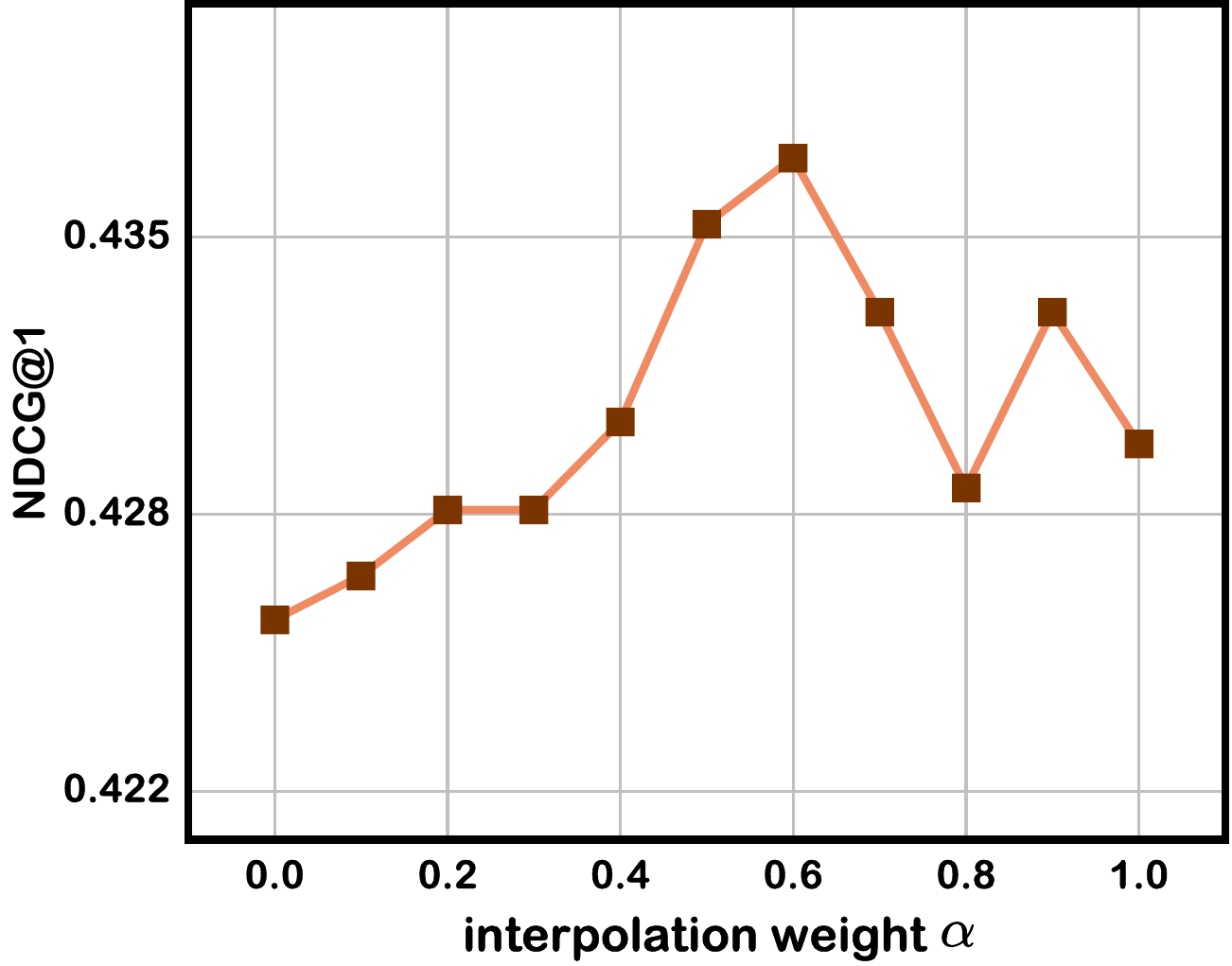}}
    \vspace{-0.2cm}
    \caption{Sensitivity analysis of the weight $\alpha$, where $M_{merged}=\alpha\cdot M_{hybrid}+(1-\alpha)\cdot M_{target}$.}
    \label{interweight} % 主图标签
\end{figure}

As shown in Figure \ref{interweight}, we investigate how changes in interpolation weight $\alpha$ affect the final model's performance under different one-to-one cross-domain scenarios. We observe that under various scenarios, Weight Average ($\alpha=0.5$) achieves sub-optimal performance with a minimal performance gap from the optimal weight. Thus Weight Average is an excellent strategy when the number of source domains increases because of search cost.

\iffalse
\subsubsection{\textbf{Why one source domain per branch?}}
In this section, we explore why each branch of \textit{WeaveRec} contains at most one source domain rather than combining multiple source domains into one branch. To investigate this, we conducted controlled experiments, fixing \textit{WeaveRec} to two branches. As shown in Figure \ref{indepth-1}, "0 Source Domain" signifies a degeneration to the target-domain LoRA, while larger values indicate mixing all source domains with the target domain in one branch.
Performance is optimal when only one source domain is mixed within a single branch. A substantial decline in model performance was observed when multiple source domains were mixed in one branch. When data from one source domain are mixed with that of the target domain to form the second branch, \textit{WeaveRec}'s performance surpasses that of target-domain LoRA, indicating the alleviation of negative transfer. However, when multiple source domains are mixed with the target domain to form the second branch, performance degradation of varying degrees occurs, suggesting negative transfer likely caused by potential gradient interference and other factors inherent in multi-task learning.
\fi

\begin{table}[!t] % table 用于创建单栏表格
    \centering % 居中表格
    \caption{Component Analysis of WeaveRec on NDCG@1.}
    \vspace{-0.2cm}
    \label{tab:component-analysis}
    \resizebox{0.4\textwidth}{!}{
    \begin{tabular}{l|ccc}
        \toprule
        \textbf{Target domain} & \textbf{Sports} & \textbf{Beauty} & \textbf{Clothing} \\
        \midrule
        Target-domain LoRA & 0.3708 & 0.4071 & 0.3643 \\
        Hybrid LoRA 1 & 0.3698 & 0.4059 & 0.3673 \\
        Hybrid LoRA 2 & 0.3786 & 0.4072 & 0.3573 \\
        Hybrid LoRA 3 & 0.3709 & 0.4029 & 0.3567 \\
        WeaveRec & \textbf{0.3897} & \textbf{0.4180} & \textbf{0.3732} \\
        \bottomrule
    \end{tabular}
    }
    \vspace{-0.4cm}
\end{table}

\begin{table}[!t] % table 用于创建单栏表格
    \centering % 居中表格
    \caption{Performance comparison of four merging methods applied to WeaveRec.}
    \vspace{-0.2cm}
    \label{tab:why-WA}
    \resizebox{0.45\textwidth}{!}{
    \begin{tabular}{l|cccc}
        \toprule
        \textbf{Setting} & \multicolumn{4}{c}{\textbf{Sports, Clothing, Food} $\to$ \textbf{Beauty}} \\
        \midrule
        \textbf{Metrics} & NDCG@1 & NDCG@3 & NDCG@5 & MRR@5 \\
        \midrule
        Target-domain Only & 0.4071 & 0.4293 & 0.4438 & 0.4314 \\
        WeaveRec w/ WA & \textbf{0.4180} & \textbf{0.4386} & \textbf{0.4543} & \textbf{0.4418}\\
        w/ LoRA-LEGO & 0.3493 & 0.3710 & 0.3872 & 0.3742 \\
        w/ Ties-Merging & 0.3466 & 0.3679 & 0.3857 & 0.3720 \\
        w/ AdaMerging & 0.4115 & 0.4318 & 0.4467 & 0.4347 \\
        \bottomrule
    \end{tabular}
    }
\end{table}

\iffalse
\begin{figure}[!t]
    \centering
    \subfloat{\includegraphics[width=0.49\columnwidth]{figures/indepth_analysis-2-a.pdf}}
    \hfil
    \subfloat{\includegraphics[width=0.49\columnwidth]{figures/indepth_analysis-2-b.pdf}}
    \vspace{-0.2cm}
    \caption{The impact of the sampling ratio on \textit{WeaveRec}'s performance. Sports is the target domain and Clothing is the source domain.}
    \label{indepth-2} % 主图标签
\end{figure}
\fi

\section{Conclusion}
In this paper, we addressed the degradation problem in cross-domain recommendation systems when applying model-merging techniques. Through experimental analysis, we identified that performance degradation occurs when source-domain models perform poorly on the target domain, leading to misleading patterns that compromise recommendation quality.
We proposed WeaveRec, a novel framework that trains a mixed-domain model and merges it with a target-domain-only model to better adapt source knowledge to target distributions. Our theoretical analysis demonstrates that WeaveRec provably reduces the upper bound of generalization error, while extensive experiments across various scenarios validate its effectiveness in consistently outperforming baseline approaches.
WeaveRec maintains the scalability advantages of model merging without additional inference costs, opening new avenues for cross-domain recommendation systems, opening new avenues for leveraging model-merging techniques in cross-domain learning.
\newpage


\bibliographystyle{ACM-Reference-Format}
\bibliography{sample-base}


\begin{thebibliography}{53}

%%% the Web address.

\ifx \showCODEN    \undefined \def \showCODEN     #1{\unskip}     \fi
\ifx \showISBNx    \undefined \def \showISBNx     #1{\unskip}     \fi
\ifx \showISBNxiii \undefined \def \showISBNxiii  #1{\unskip}     \fi
\ifx \showISSN     \undefined \def \showISSN      #1{\unskip}     \fi
\ifx \showLCCN     \undefined \def \showLCCN      #1{\unskip}     \fi
\ifx \shownote     \undefined \def \shownote      #1{#1}          \fi
\ifx \showarticletitle \undefined \def \showarticletitle #1{#1}   \fi
\ifx \showURL      \undefined \def \showURL       {\relax}        \fi
% invisible to TeX
\providecommand\bibfield[2]{#2}
\providecommand\bibinfo[2]{#2}
\providecommand\natexlab[1]{#1}
\providecommand\showeprint[2][]{arXiv:#2}

\bibitem[Bao et~al\mbox{.}(2025)]%
        {bao2025heterogeneoususermodelingllmbased}
\bibfield{author}{\bibinfo{person}{Honghui Bao}, \bibinfo{person}{Wenjie Wang}, \bibinfo{person}{Xinyu Lin}, \bibinfo{person}{Fengbin Zhu}, \bibinfo{person}{Teng Sun}, \bibinfo{person}{Fuli Feng}, {and} \bibinfo{person}{Tat-Seng Chua}.} \bibinfo{year}{2025}\natexlab{}.
\newblock \showarticletitle{Heterogeneous User Modeling for LLM-based Recommendation}. In \bibinfo{booktitle}{\emph{RecSys}}.
\newblock


\bibitem[Bao et~al\mbox{.}(2023)]%
        {tallrec}
\bibfield{author}{\bibinfo{person}{Keqin Bao}, \bibinfo{person}{Jizhi Zhang}, \bibinfo{person}{Yang Zhang}, \bibinfo{person}{Wenjie Wang}, \bibinfo{person}{Fuli Feng}, {and} \bibinfo{person}{Xiangnan He}.} \bibinfo{year}{2023}\natexlab{}.
\newblock \showarticletitle{Tallrec: An effective and efficient tuning framework to align large language model with recommendation}. In \bibinfo{booktitle}{\emph{Proceedings of the 17th ACM conference on recommender systems}}. \bibinfo{pages}{1007--1014}.
\newblock


\bibitem[Ben-David et~al\mbox{.}(2010)]%
        {10.1007/s10994-009-5152-4}
\bibfield{author}{\bibinfo{person}{Shai Ben-David}, \bibinfo{person}{John Blitzer}, \bibinfo{person}{Koby Crammer}, \bibinfo{person}{Alex Kulesza}, \bibinfo{person}{Fernando Pereira}, {and} \bibinfo{person}{Jennifer~Wortman Vaughan}.} \bibinfo{year}{2010}\natexlab{}.
\newblock \showarticletitle{A theory of learning from different domains}.
\newblock \bibinfo{journal}{\emph{Mach. Learn.}} \bibinfo{volume}{79}, \bibinfo{number}{1–2} (\bibinfo{date}{May} \bibinfo{year}{2010}), \bibinfo{pages}{151–175}.
\newblock
\showISSN{0885-6125}
\href{https://doi.org/10.1007/s10994-009-5152-4}{doi:\nolinkurl{10.1007/s10994-009-5152-4}}


\bibitem[Biggs et~al\mbox{.}(2024)]%
        {diffusionsoup}
\bibfield{author}{\bibinfo{person}{Benjamin Biggs}, \bibinfo{person}{Arjun Seshadri}, \bibinfo{person}{Yang Zou}, \bibinfo{person}{Achin Jain}, \bibinfo{person}{Aditya Golatkar}, \bibinfo{person}{Yusheng Xie}, \bibinfo{person}{Alessandro Achille}, \bibinfo{person}{Ashwin Swaminathan}, {and} \bibinfo{person}{Stefano Soatto}.} \bibinfo{year}{2024}\natexlab{}.
\newblock \showarticletitle{Diffusion soup: Model merging for text-to-image diffusion models}. In \bibinfo{booktitle}{\emph{European Conference on Computer Vision}}. Springer, \bibinfo{pages}{257--274}.
\newblock


\bibitem[Cao et~al\mbox{.}(2023)]%
        {10.1145/3539597.3570366}
\bibfield{author}{\bibinfo{person}{Jiangxia Cao}, \bibinfo{person}{Shaoshuai Li}, \bibinfo{person}{Bowen Yu}, \bibinfo{person}{Xiaobo Guo}, \bibinfo{person}{Tingwen Liu}, {and} \bibinfo{person}{Bin Wang}.} \bibinfo{year}{2023}\natexlab{}.
\newblock \showarticletitle{Towards Universal Cross-Domain Recommendation}. In \bibinfo{booktitle}{\emph{Proceedings of the Sixteenth ACM International Conference on Web Search and Data Mining}} (Singapore, Singapore) \emph{(\bibinfo{series}{WSDM '23})}. \bibinfo{publisher}{Association for Computing Machinery}, \bibinfo{address}{New York, NY, USA}, \bibinfo{pages}{78–86}.
\newblock
\showISBNx{9781450394079}
\href{https://doi.org/10.1145/3539597.3570366}{doi:\nolinkurl{10.1145/3539597.3570366}}


\bibitem[Cui et~al\mbox{.}(2022)]%
        {m6rec}
\bibfield{author}{\bibinfo{person}{Zeyu Cui}, \bibinfo{person}{Jianxin Ma}, \bibinfo{person}{Chang Zhou}, \bibinfo{person}{Jingren Zhou}, {and} \bibinfo{person}{Hongxia Yang}.} \bibinfo{year}{2022}\natexlab{}.
\newblock \bibinfo{title}{M6-Rec: Generative Pretrained Language Models are Open-Ended Recommender Systems}.
\newblock
\showeprint[arxiv]{2205.08084}~[cs.IR]
\urldef\tempurl%
\url{https://arxiv.org/abs/2205.08084}
\showURL{%
\tempurl}


\bibitem[Dai et~al\mbox{.}(2023)]%
        {10.1145/3604915.3610646}
\bibfield{author}{\bibinfo{person}{Sunhao Dai}, \bibinfo{person}{Ninglu Shao}, \bibinfo{person}{Haiyuan Zhao}, \bibinfo{person}{Weijie Yu}, \bibinfo{person}{Zihua Si}, \bibinfo{person}{Chen Xu}, \bibinfo{person}{Zhongxiang Sun}, \bibinfo{person}{Xiao Zhang}, {and} \bibinfo{person}{Jun Xu}.} \bibinfo{year}{2023}\natexlab{}.
\newblock \showarticletitle{Uncovering ChatGPT’s Capabilities in Recommender Systems}. In \bibinfo{booktitle}{\emph{Proceedings of the 17th ACM Conference on Recommender Systems}} (Singapore, Singapore) \emph{(\bibinfo{series}{RecSys '23})}. \bibinfo{publisher}{Association for Computing Machinery}, \bibinfo{address}{New York, NY, USA}, \bibinfo{pages}{1126–1132}.
\newblock
\showISBNx{9798400702419}
\href{https://doi.org/10.1145/3604915.3610646}{doi:\nolinkurl{10.1145/3604915.3610646}}


\bibitem[Dong et~al\mbox{.}(2024)]%
        {dong2024survey}
\bibfield{author}{\bibinfo{person}{Qingxiu Dong}, \bibinfo{person}{Lei Li}, \bibinfo{person}{Damai Dai}, \bibinfo{person}{Ce Zheng}, \bibinfo{person}{Jingyuan Ma}, \bibinfo{person}{Rui Li}, \bibinfo{person}{Heming Xia}, \bibinfo{person}{Jingjing Xu}, \bibinfo{person}{Zhiyong Wu}, \bibinfo{person}{Baobao Chang}, {et~al\mbox{.}}} \bibinfo{year}{2024}\natexlab{}.
\newblock \showarticletitle{A survey on in-context learning}. In \bibinfo{booktitle}{\emph{Proceedings of the 2024 Conference on Empirical Methods in Natural Language Processing}}. \bibinfo{pages}{1107--1128}.
\newblock


\bibitem[Dong et~al\mbox{.}(2020)]%
        {dong2020survey}
\bibfield{author}{\bibinfo{person}{Xibin Dong}, \bibinfo{person}{Zhiwen Yu}, \bibinfo{person}{Wenming Cao}, \bibinfo{person}{Yifan Shi}, {and} \bibinfo{person}{Qianli Ma}.} \bibinfo{year}{2020}\natexlab{}.
\newblock \showarticletitle{A survey on ensemble learning}.
\newblock \bibinfo{journal}{\emph{Frontiers of Computer Science}} \bibinfo{volume}{14}, \bibinfo{number}{2} (\bibinfo{year}{2020}), \bibinfo{pages}{241--258}.
\newblock
\href{https://doi.org/10.1007/s11704-019-8208-z}{doi:\nolinkurl{10.1007/s11704-019-8208-z}}


\bibitem[Frankle et~al\mbox{.}(2020)]%
        {frankle2020linear}
\bibfield{author}{\bibinfo{person}{Jonathan Frankle}, \bibinfo{person}{Gintare~Karolina Dziugaite}, \bibinfo{person}{Daniel~M Roy}, {and} \bibinfo{person}{Michael Carbin}.} \bibinfo{year}{2020}\natexlab{}.
\newblock \showarticletitle{Linear Mode Connectivity and the Lottery Ticket Hypothesis}. In \bibinfo{booktitle}{\emph{Proceedings of the 37th International Conference on Machine Learning}} \emph{(\bibinfo{series}{Proceedings of Machine Learning Research}, Vol.~\bibinfo{volume}{119})}. \bibinfo{publisher}{PMLR}, \bibinfo{pages}{3259--3269}.
\newblock


\bibitem[Garipov et~al\mbox{.}(2018)]%
        {10.5555/3327546.3327556}
\bibfield{author}{\bibinfo{person}{Timur Garipov}, \bibinfo{person}{Pavel Izmailov}, \bibinfo{person}{Dmitrii Podoprikhin}, \bibinfo{person}{Dmitry Vetrov}, {and} \bibinfo{person}{Andrew~Gordon Wilson}.} \bibinfo{year}{2018}\natexlab{}.
\newblock \showarticletitle{Loss surfaces, mode connectivity, and fast ensembling of DNNs}. In \bibinfo{booktitle}{\emph{Proceedings of the 32nd International Conference on Neural Information Processing Systems}} (Montr\'{e}al, Canada) \emph{(\bibinfo{series}{NIPS'18})}. \bibinfo{publisher}{Curran Associates Inc.}, \bibinfo{address}{Red Hook, NY, USA}, \bibinfo{pages}{8803–8812}.
\newblock


\bibitem[Geng et~al\mbox{.}(2022a)]%
        {p5}
\bibfield{author}{\bibinfo{person}{Shijie Geng}, \bibinfo{person}{Shuchang Liu}, \bibinfo{person}{Zuohui Fu}, \bibinfo{person}{Yingqiang Ge}, {and} \bibinfo{person}{Yongfeng Zhang}.} \bibinfo{year}{2022}\natexlab{a}.
\newblock \showarticletitle{Recommendation as language processing (rlp): A unified pretrain, personalized prompt \& predict paradigm (p5)}. In \bibinfo{booktitle}{\emph{Proceedings of the 16th ACM conference on recommender systems}}. \bibinfo{pages}{299--315}.
\newblock


\bibitem[Geng et~al\mbox{.}(2022b)]%
        {10.1145/3523227.3546767}
\bibfield{author}{\bibinfo{person}{Shijie Geng}, \bibinfo{person}{Shuchang Liu}, \bibinfo{person}{Zuohui Fu}, \bibinfo{person}{Yingqiang Ge}, {and} \bibinfo{person}{Yongfeng Zhang}.} \bibinfo{year}{2022}\natexlab{b}.
\newblock \showarticletitle{Recommendation as Language Processing (RLP): A Unified Pretrain, Personalized Prompt \& Predict Paradigm (P5)}. In \bibinfo{booktitle}{\emph{Proceedings of the 16th ACM Conference on Recommender Systems}} (Seattle, WA, USA) \emph{(\bibinfo{series}{RecSys '22})}. \bibinfo{publisher}{Association for Computing Machinery}, \bibinfo{address}{New York, NY, USA}, \bibinfo{pages}{299–315}.
\newblock
\showISBNx{9781450392785}
\href{https://doi.org/10.1145/3523227.3546767}{doi:\nolinkurl{10.1145/3523227.3546767}}


\bibitem[Guo et~al\mbox{.}(2023a)]%
        {PLCR}
\bibfield{author}{\bibinfo{person}{Lei Guo}, \bibinfo{person}{Chunxiao Wang}, \bibinfo{person}{Xinhua Wang}, \bibinfo{person}{Lei Zhu}, {and} \bibinfo{person}{Hongzhi Yin}.} \bibinfo{year}{2023}\natexlab{a}.
\newblock \showarticletitle{Automated prompting for non-overlapping cross-domain sequential recommendation}.
\newblock \bibinfo{journal}{\emph{arXiv preprint arXiv:2304.04218}} (\bibinfo{year}{2023}).
\newblock


\bibitem[Guo et~al\mbox{.}(2023b)]%
        {10.1109/TKDE.2022.3185101}
\bibfield{author}{\bibinfo{person}{Lei Guo}, \bibinfo{person}{Jinyu Zhang}, \bibinfo{person}{Tong Chen}, \bibinfo{person}{Xinhua Wang}, {and} \bibinfo{person}{Hongzhi Yin}.} \bibinfo{year}{2023}\natexlab{b}.
\newblock \showarticletitle{Reinforcement Learning-Enhanced Shared-Account Cross-Domain Sequential Recommendation}.
\newblock \bibinfo{journal}{\emph{IEEE Trans. on Knowl. and Data Eng.}} \bibinfo{volume}{35}, \bibinfo{number}{7} (\bibinfo{date}{July} \bibinfo{year}{2023}), \bibinfo{pages}{7397–7411}.
\newblock
\showISSN{1041-4347}
\href{https://doi.org/10.1109/TKDE.2022.3185101}{doi:\nolinkurl{10.1109/TKDE.2022.3185101}}


\bibitem[Hidasi et~al\mbox{.}(2016)]%
        {gru4rec}
\bibfield{author}{\bibinfo{person}{Bal{\'{a}}zs Hidasi}, \bibinfo{person}{Alexandros Karatzoglou}, \bibinfo{person}{Linas Baltrunas}, {and} \bibinfo{person}{Domonkos Tikk}.} \bibinfo{year}{2016}\natexlab{}.
\newblock \showarticletitle{Session-based Recommendations with Recurrent Neural Networks}. In \bibinfo{booktitle}{\emph{4th International Conference on Learning Representations, {ICLR} 2016, San Juan, Puerto Rico, May 2-4, 2016, Conference Track Proceedings}}, \bibfield{editor}{\bibinfo{person}{Yoshua Bengio} {and} \bibinfo{person}{Yann LeCun}} (Eds.).
\newblock
\urldef\tempurl%
\url{http://arxiv.org/abs/1511.06939}
\showURL{%
\tempurl}


\bibitem[Hou et~al\mbox{.}(2023)]%
        {hou2023vqrec}
\bibfield{author}{\bibinfo{person}{Yupeng Hou}, \bibinfo{person}{Zhankui He}, \bibinfo{person}{Julian McAuley}, {and} \bibinfo{person}{Wayne~Xin Zhao}.} \bibinfo{year}{2023}\natexlab{}.
\newblock \showarticletitle{Learning Vector-Quantized Item Representation for Transferable Sequential Recommenders}. In \bibinfo{booktitle}{\emph{{TheWebConf}}}.
\newblock


\bibitem[Hou et~al\mbox{.}(2022)]%
        {unisrec}
\bibfield{author}{\bibinfo{person}{Yupeng Hou}, \bibinfo{person}{Shanlei Mu}, \bibinfo{person}{Wayne~Xin Zhao}, \bibinfo{person}{Yaliang Li}, \bibinfo{person}{Bolin Ding}, {and} \bibinfo{person}{Ji-Rong Wen}.} \bibinfo{year}{2022}\natexlab{}.
\newblock \showarticletitle{Towards Universal Sequence Representation Learning for Recommender Systems}. In \bibinfo{booktitle}{\emph{Proceedings of the 28th ACM SIGKDD Conference on Knowledge Discovery and Data Mining}} (Washington DC, USA) \emph{(\bibinfo{series}{KDD '22})}. \bibinfo{publisher}{Association for Computing Machinery}, \bibinfo{address}{New York, NY, USA}, \bibinfo{pages}{585–593}.
\newblock
\showISBNx{9781450393850}
\href{https://doi.org/10.1145/3534678.3539381}{doi:\nolinkurl{10.1145/3534678.3539381}}


\bibitem[Hu et~al\mbox{.}(2022)]%
        {2022lora}
\bibfield{author}{\bibinfo{person}{Edward~J Hu}, \bibinfo{person}{Yelong Shen}, \bibinfo{person}{Phillip Wallis}, \bibinfo{person}{Zeyuan Allen-Zhu}, \bibinfo{person}{Yuanzhi Li}, \bibinfo{person}{Shean Wang}, \bibinfo{person}{Lu Wang}, \bibinfo{person}{Weizhu Chen}, {et~al\mbox{.}}} \bibinfo{year}{2022}\natexlab{}.
\newblock \showarticletitle{Lora: Low-rank adaptation of large language models.}
\newblock \bibinfo{journal}{\emph{ICLR}} \bibinfo{volume}{1}, \bibinfo{number}{2} (\bibinfo{year}{2022}), \bibinfo{pages}{3}.
\newblock


\bibitem[Hu et~al\mbox{.}(2024)]%
        {ext-sub}
\bibfield{author}{\bibinfo{person}{Xinshuo Hu}, \bibinfo{person}{Dongfang Li}, \bibinfo{person}{Baotian Hu}, \bibinfo{person}{Zihao Zheng}, \bibinfo{person}{Zhenyu Liu}, {and} \bibinfo{person}{Min Zhang}.} \bibinfo{year}{2024}\natexlab{}.
\newblock \showarticletitle{Separate the wheat from the chaff: model deficiency unlearning via parameter-efficient module operation}. In \bibinfo{booktitle}{\emph{Proceedings of the Thirty-Eighth AAAI Conference on Artificial Intelligence and Thirty-Sixth Conference on Innovative Applications of Artificial Intelligence and Fourteenth Symposium on Educational Advances in Artificial Intelligence}} \emph{(\bibinfo{series}{AAAI'24/IAAI'24/EAAI'24})}. \bibinfo{publisher}{AAAI Press}, Article \bibinfo{articleno}{2036}, \bibinfo{numpages}{9}~pages.
\newblock
\showISBNx{978-1-57735-887-9}
\href{https://doi.org/10.1609/aaai.v38i16.29784}{doi:\nolinkurl{10.1609/aaai.v38i16.29784}}


\bibitem[Huang and Chang(2023)]%
        {huang-chang-2023-towards}
\bibfield{author}{\bibinfo{person}{Jie Huang} {and} \bibinfo{person}{Kevin Chen-Chuan Chang}.} \bibinfo{year}{2023}\natexlab{}.
\newblock \showarticletitle{Towards Reasoning in Large Language Models: A Survey}. In \bibinfo{booktitle}{\emph{Findings of the Association for Computational Linguistics: ACL 2023}}, \bibfield{editor}{\bibinfo{person}{Anna Rogers}, \bibinfo{person}{Jordan Boyd-Graber}, {and} \bibinfo{person}{Naoaki Okazaki}} (Eds.). \bibinfo{publisher}{Association for Computational Linguistics}, \bibinfo{address}{Toronto, Canada}, \bibinfo{pages}{1049--1065}.
\newblock
\href{https://doi.org/10.18653/v1/2023.findings-acl.67}{doi:\nolinkurl{10.18653/v1/2023.findings-acl.67}}


\bibitem[Ilharco et~al\mbox{.}(2023)]%
        {taskarithmetic}
\bibfield{author}{\bibinfo{person}{Gabriel Ilharco}, \bibinfo{person}{Marco~Tulio Ribeiro}, \bibinfo{person}{Mitchell Wortsman}, \bibinfo{person}{Suchin Gururangan}, \bibinfo{person}{Ludwig Schmidt}, \bibinfo{person}{Hannaneh Hajishirzi}, {and} \bibinfo{person}{Ali Farhadi}.} \bibinfo{year}{2023}\natexlab{}.
\newblock \bibinfo{title}{Editing Models with Task Arithmetic}.
\newblock
\showeprint[arxiv]{2212.04089}~[cs.LG]
\urldef\tempurl%
\url{https://arxiv.org/abs/2212.04089}
\showURL{%
\tempurl}


\bibitem[Izmailov et~al\mbox{.}(2018)]%
        {izmailov2018averaging}
\bibfield{author}{\bibinfo{person}{Pavel Izmailov}, \bibinfo{person}{Dmitrii Podoprikhin}, \bibinfo{person}{Timur Garipov}, \bibinfo{person}{Dmitry Vetrov}, {and} \bibinfo{person}{Andrew~Gordon Wilson}.} \bibinfo{year}{2018}\natexlab{}.
\newblock \showarticletitle{Averaging weights leads to wider optima and better generalization}.
\newblock \bibinfo{journal}{\emph{arXiv preprint arXiv:1803.05407}} (\bibinfo{year}{2018}).
\newblock


\bibitem[Kang and McAuley(2018)]%
        {sasrec}
\bibfield{author}{\bibinfo{person}{Wang-Cheng Kang} {and} \bibinfo{person}{Julian McAuley}.} \bibinfo{year}{2018}\natexlab{}.
\newblock \showarticletitle{Self-attentive sequential recommendation}. In \bibinfo{booktitle}{\emph{2018 IEEE international conference on data mining (ICDM)}}. IEEE, \bibinfo{pages}{197--206}.
\newblock


\bibitem[Li et~al\mbox{.}(2023)]%
        {recformer}
\bibfield{author}{\bibinfo{person}{Jiacheng Li}, \bibinfo{person}{Ming Wang}, \bibinfo{person}{Jin Li}, \bibinfo{person}{Jinmiao Fu}, \bibinfo{person}{Xin Shen}, \bibinfo{person}{Jingbo Shang}, {and} \bibinfo{person}{Julian McAuley}.} \bibinfo{year}{2023}\natexlab{}.
\newblock \showarticletitle{Text is all you need: Learning language representations for sequential recommendation}. In \bibinfo{booktitle}{\emph{Proceedings of the 29th ACM SIGKDD Conference on Knowledge Discovery and Data Mining}}. \bibinfo{pages}{1258--1267}.
\newblock


\bibitem[Lin et~al\mbox{.}(2025)]%
        {10.1145/3678004}
\bibfield{author}{\bibinfo{person}{Jianghao Lin}, \bibinfo{person}{Xinyi Dai}, \bibinfo{person}{Yunjia Xi}, \bibinfo{person}{Weiwen Liu}, \bibinfo{person}{Bo Chen}, \bibinfo{person}{Hao Zhang}, \bibinfo{person}{Yong Liu}, \bibinfo{person}{Chuhan Wu}, \bibinfo{person}{Xiangyang Li}, \bibinfo{person}{Chenxu Zhu}, \bibinfo{person}{Huifeng Guo}, \bibinfo{person}{Yong Yu}, \bibinfo{person}{Ruiming Tang}, {and} \bibinfo{person}{Weinan Zhang}.} \bibinfo{year}{2025}\natexlab{}.
\newblock \showarticletitle{How Can Recommender Systems Benefit from Large Language Models: A Survey}.
\newblock \bibinfo{journal}{\emph{ACM Trans. Inf. Syst.}} \bibinfo{volume}{43}, \bibinfo{number}{2}, Article \bibinfo{articleno}{28} (\bibinfo{date}{Jan.} \bibinfo{year}{2025}), \bibinfo{numpages}{47}~pages.
\newblock
\showISSN{1046-8188}
\href{https://doi.org/10.1145/3678004}{doi:\nolinkurl{10.1145/3678004}}


\bibitem[Lin et~al\mbox{.}(2024)]%
        {2024bridging}
\bibfield{author}{\bibinfo{person}{Xinyu Lin}, \bibinfo{person}{Wenjie Wang}, \bibinfo{person}{Yongqi Li}, \bibinfo{person}{Fuli Feng}, \bibinfo{person}{See-Kiong Ng}, {and} \bibinfo{person}{Tat-Seng Chua}.} \bibinfo{year}{2024}\natexlab{}.
\newblock \showarticletitle{Bridging items and language: A transition paradigm for large language model-based recommendation}. In \bibinfo{booktitle}{\emph{Proceedings of the 30th ACM SIGKDD Conference on Knowledge Discovery and Data Mining}}. \bibinfo{pages}{1816--1826}.
\newblock


\bibitem[Liu et~al\mbox{.}(2024a)]%
        {mcrpl}
\bibfield{author}{\bibinfo{person}{Hao Liu}, \bibinfo{person}{Lei Guo}, \bibinfo{person}{Lei Zhu}, \bibinfo{person}{Yongqiang Jiang}, \bibinfo{person}{Min Gao}, {and} \bibinfo{person}{Hongzhi Yin}.} \bibinfo{year}{2024}\natexlab{a}.
\newblock \showarticletitle{MCRPL: A Pretrain, prompt, and fine-tune paradigm for non-overlapping many-to-one cross-domain recommendation}.
\newblock \bibinfo{journal}{\emph{ACM Transactions on Information Systems}} \bibinfo{volume}{42}, \bibinfo{number}{4} (\bibinfo{year}{2024}), \bibinfo{pages}{1--24}.
\newblock


\bibitem[Liu et~al\mbox{.}(2024b)]%
        {10.1609/aaai.v38i8.28723}
\bibfield{author}{\bibinfo{person}{Jing Liu}, \bibinfo{person}{Lele Sun}, \bibinfo{person}{Weizhi Nie}, \bibinfo{person}{Peiguang Jing}, {and} \bibinfo{person}{Yuting Su}.} \bibinfo{year}{2024}\natexlab{b}.
\newblock \showarticletitle{Graph disentangled contrastive learning with personalized transfer for cross-domain recommendation}. In \bibinfo{booktitle}{\emph{Proceedings of the Thirty-Eighth AAAI Conference on Artificial Intelligence and Thirty-Sixth Conference on Innovative Applications of Artificial Intelligence and Fourteenth Symposium on Educational Advances in Artificial Intelligence}} \emph{(\bibinfo{series}{AAAI'24/IAAI'24/EAAI'24})}. \bibinfo{publisher}{AAAI Press}, Article \bibinfo{articleno}{975}, \bibinfo{numpages}{9}~pages.
\newblock
\showISBNx{978-1-57735-887-9}
\href{https://doi.org/10.1609/aaai.v38i8.28723}{doi:\nolinkurl{10.1609/aaai.v38i8.28723}}


\bibitem[Liu et~al\mbox{.}(2025)]%
        {LLM4CDR}
\bibfield{author}{\bibinfo{person}{Xinyi Liu}, \bibinfo{person}{Ruijie Wang}, \bibinfo{person}{Dachun Sun}, \bibinfo{person}{Dilek Hakkani~Tur}, {and} \bibinfo{person}{Tarek Abdelzaher}.} \bibinfo{year}{2025}\natexlab{}.
\newblock \showarticletitle{Uncovering Cross-Domain Recommendation Ability of Large Language Models}. In \bibinfo{booktitle}{\emph{Companion Proceedings of the ACM on Web Conference 2025}} (Sydney NSW, Australia) \emph{(\bibinfo{series}{WWW '25})}. \bibinfo{publisher}{Association for Computing Machinery}, \bibinfo{address}{New York, NY, USA}, \bibinfo{pages}{2736–2743}.
\newblock
\showISBNx{9798400713316}
\href{https://doi.org/10.1145/3701716.3717850}{doi:\nolinkurl{10.1145/3701716.3717850}}


\bibitem[Peng et~al\mbox{.}(2024a)]%
        {ecellm}
\bibfield{author}{\bibinfo{person}{Bo Peng}, \bibinfo{person}{Xinyi Ling}, \bibinfo{person}{Ziru Chen}, \bibinfo{person}{Huan Sun}, {and} \bibinfo{person}{Xia Ning}.} \bibinfo{year}{2024}\natexlab{a}.
\newblock \showarticletitle{eCe{LLM}: Generalizing Large Language Models for E-commerce from Large-scale, High-quality Instruction Data}. In \bibinfo{booktitle}{\emph{Forty-first International Conference on Machine Learning}}.
\newblock
\urldef\tempurl%
\url{https://openreview.net/forum?id=LWRI4uPG2X}
\showURL{%
\tempurl}


\bibitem[Peng et~al\mbox{.}(2024b)]%
        {10.5555/3692070.3693702}
\bibfield{author}{\bibinfo{person}{Bo Peng}, \bibinfo{person}{Xinyi Ling}, \bibinfo{person}{Ziru Chen}, \bibinfo{person}{Huan Sun}, {and} \bibinfo{person}{Xia Ning}.} \bibinfo{year}{2024}\natexlab{b}.
\newblock \showarticletitle{eCeLLM: generalizing large language models for E-commerce from large-scale, high-quality instruction data}. In \bibinfo{booktitle}{\emph{Proceedings of the 41st International Conference on Machine Learning}} (Vienna, Austria) \emph{(\bibinfo{series}{ICML'24})}. \bibinfo{publisher}{JMLR.org}, Article \bibinfo{articleno}{1632}, \bibinfo{numpages}{43}~pages.
\newblock


\bibitem[Sanner et~al\mbox{.}(2023)]%
        {10.1145/3604915.3608845}
\bibfield{author}{\bibinfo{person}{Scott Sanner}, \bibinfo{person}{Krisztian Balog}, \bibinfo{person}{Filip Radlinski}, \bibinfo{person}{Ben Wedin}, {and} \bibinfo{person}{Lucas Dixon}.} \bibinfo{year}{2023}\natexlab{}.
\newblock \showarticletitle{Large Language Models are Competitive Near Cold-start Recommenders for Language- and Item-based Preferences}. In \bibinfo{booktitle}{\emph{Proceedings of the 17th ACM Conference on Recommender Systems}} (Singapore, Singapore) \emph{(\bibinfo{series}{RecSys '23})}. \bibinfo{publisher}{Association for Computing Machinery}, \bibinfo{address}{New York, NY, USA}, \bibinfo{pages}{890–896}.
\newblock
\showISBNx{9798400702419}
\href{https://doi.org/10.1145/3604915.3608845}{doi:\nolinkurl{10.1145/3604915.3608845}}


\bibitem[Sun et~al\mbox{.}(2019)]%
        {bert4rec}
\bibfield{author}{\bibinfo{person}{Fei Sun}, \bibinfo{person}{Jun Liu}, \bibinfo{person}{Jian Wu}, \bibinfo{person}{Changhua Pei}, \bibinfo{person}{Xiao Lin}, \bibinfo{person}{Wenwu Ou}, {and} \bibinfo{person}{Peng Jiang}.} \bibinfo{year}{2019}\natexlab{}.
\newblock \showarticletitle{BERT4Rec: Sequential Recommendation with Bidirectional Encoder Representations from Transformer}. In \bibinfo{booktitle}{\emph{Proceedings of the 28th ACM International Conference on Information and Knowledge Management}} (Beijing, China) \emph{(\bibinfo{series}{CIKM '19})}. \bibinfo{publisher}{Association for Computing Machinery}, \bibinfo{address}{New York, NY, USA}, \bibinfo{pages}{1441–1450}.
\newblock
\showISBNx{9781450369763}
\href{https://doi.org/10.1145/3357384.3357895}{doi:\nolinkurl{10.1145/3357384.3357895}}


\bibitem[Tang et~al\mbox{.}(2025)]%
        {one-model-for-all}
\bibfield{author}{\bibinfo{person}{Zuoli Tang}, \bibinfo{person}{Zhaoxin Huan}, \bibinfo{person}{Zihao Li}, \bibinfo{person}{Xiaolu Zhang}, \bibinfo{person}{Jun Hu}, \bibinfo{person}{Chilin Fu}, \bibinfo{person}{Jun Zhou}, \bibinfo{person}{Lixin Zou}, {and} \bibinfo{person}{Chenliang Li}.} \bibinfo{year}{2025}\natexlab{}.
\newblock \showarticletitle{One Model for All: Large Language Models Are Domain-Agnostic Recommendation Systems}.
\newblock \bibinfo{journal}{\emph{ACM Trans. Inf. Syst.}} \bibinfo{volume}{43}, \bibinfo{number}{5}, Article \bibinfo{articleno}{118} (\bibinfo{date}{July} \bibinfo{year}{2025}), \bibinfo{numpages}{27}~pages.
\newblock
\showISSN{1046-8188}
\href{https://doi.org/10.1145/3705727}{doi:\nolinkurl{10.1145/3705727}}


\bibitem[Utans(1996)]%
        {1996Weight}
\bibfield{author}{\bibinfo{person}{Joachim Utans}.} \bibinfo{year}{1996}\natexlab{}.
\newblock \showarticletitle{Weight Averaging for Neural Networks and Local Resampling Schemes}. In \bibinfo{booktitle}{\emph{AAAI-96 Workshop on Integrating Multiple Learned Models.}}
\newblock


\bibitem[Wang et~al\mbox{.}(2023)]%
        {wang2023rethinking}
\bibfield{author}{\bibinfo{person}{Xiaolei Wang}, \bibinfo{person}{Xinyu Tang}, \bibinfo{person}{Xin Zhao}, \bibinfo{person}{Jingyuan Wang}, {and} \bibinfo{person}{Ji-Rong Wen}.} \bibinfo{year}{2023}\natexlab{}.
\newblock \showarticletitle{Rethinking the Evaluation for Conversational Recommendation in the Era of Large Language Models}. In \bibinfo{booktitle}{\emph{The 2023 Conference on Empirical Methods in Natural Language Processing}}.
\newblock
\urldef\tempurl%
\url{https://openreview.net/forum?id=O4kDO3yS9B}
\showURL{%
\tempurl}


\bibitem[Wang et~al\mbox{.}(2024)]%
        {10.1109/TKDE.2024.3511602}
\bibfield{author}{\bibinfo{person}{Zihan Wang}, \bibinfo{person}{Yonghui Yang}, \bibinfo{person}{Le Wu}, \bibinfo{person}{Richang Hong}, {and} \bibinfo{person}{Meng Wang}.} \bibinfo{year}{2024}\natexlab{}.
\newblock \showarticletitle{Making Non-Overlapping Matters: An Unsupervised Alignment Enhanced Cross-Domain Cold-Start Recommendation}.
\newblock \bibinfo{journal}{\emph{IEEE Trans. on Knowl. and Data Eng.}} \bibinfo{volume}{37}, \bibinfo{number}{4} (\bibinfo{date}{Dec.} \bibinfo{year}{2024}), \bibinfo{pages}{2001–2014}.
\newblock
\showISSN{1041-4347}
\href{https://doi.org/10.1109/TKDE.2024.3511602}{doi:\nolinkurl{10.1109/TKDE.2024.3511602}}


\bibitem[Wortsman et~al\mbox{.}(2022a)]%
        {wortsman2022model}
\bibfield{author}{\bibinfo{person}{Mitchell Wortsman}, \bibinfo{person}{Gabriel Ilharco}, \bibinfo{person}{Samir~Ya Gadre}, \bibinfo{person}{Rebecca Roelofs}, \bibinfo{person}{Raphael Gontijo-Lopes}, \bibinfo{person}{Ari~S Morcos}, \bibinfo{person}{Hongseok Namkoong}, \bibinfo{person}{Ali Farhadi}, \bibinfo{person}{Yair Carmon}, \bibinfo{person}{Simon Kornblith}, {and} \bibinfo{person}{Ludwig Schmidt}.} \bibinfo{year}{2022}\natexlab{a}.
\newblock \showarticletitle{Model soups: averaging weights of multiple fine-tuned models improves accuracy without increasing inference time}. In \bibinfo{booktitle}{\emph{Proceedings of the 39th International Conference on Machine Learning}} \emph{(\bibinfo{series}{Proceedings of Machine Learning Research}, Vol.~\bibinfo{volume}{162})}. \bibinfo{publisher}{PMLR}, \bibinfo{pages}{23965--23998}.
\newblock


\bibitem[Wortsman et~al\mbox{.}(2022b)]%
        {pmlr-v162-wortsman22a}
\bibfield{author}{\bibinfo{person}{Mitchell Wortsman}, \bibinfo{person}{Gabriel Ilharco}, \bibinfo{person}{Samir~Ya Gadre}, \bibinfo{person}{Rebecca Roelofs}, \bibinfo{person}{Raphael Gontijo-Lopes}, \bibinfo{person}{Ari~S Morcos}, \bibinfo{person}{Hongseok Namkoong}, \bibinfo{person}{Ali Farhadi}, \bibinfo{person}{Yair Carmon}, \bibinfo{person}{Simon Kornblith}, {and} \bibinfo{person}{Ludwig Schmidt}.} \bibinfo{year}{2022}\natexlab{b}.
\newblock \showarticletitle{Model soups: averaging weights of multiple fine-tuned models improves accuracy without increasing inference time}. In \bibinfo{booktitle}{\emph{Proceedings of the 39th International Conference on Machine Learning}} \emph{(\bibinfo{series}{Proceedings of Machine Learning Research}, Vol.~\bibinfo{volume}{162})}, \bibfield{editor}{\bibinfo{person}{Kamalika Chaudhuri}, \bibinfo{person}{Stefanie Jegelka}, \bibinfo{person}{Le~Song}, \bibinfo{person}{Csaba Szepesvari}, \bibinfo{person}{Gang Niu}, {and} \bibinfo{person}{Sivan Sabato}} (Eds.). \bibinfo{publisher}{PMLR}, \bibinfo{pages}{23965--23998}.
\newblock
\urldef\tempurl%
\url{https://proceedings.mlr.press/v162/wortsman22a.html}
\showURL{%
\tempurl}


\bibitem[Wu et~al\mbox{.}(2024)]%
        {10.1007/s11280-024-01291-2}
\bibfield{author}{\bibinfo{person}{Likang Wu}, \bibinfo{person}{Zhi Zheng}, \bibinfo{person}{Zhaopeng Qiu}, \bibinfo{person}{Hao Wang}, \bibinfo{person}{Hongchao Gu}, \bibinfo{person}{Tingjia Shen}, \bibinfo{person}{Chuan Qin}, \bibinfo{person}{Chen Zhu}, \bibinfo{person}{Hengshu Zhu}, \bibinfo{person}{Qi Liu}, \bibinfo{person}{Hui Xiong}, {and} \bibinfo{person}{Enhong Chen}.} \bibinfo{year}{2024}\natexlab{}.
\newblock \showarticletitle{A survey on large language models for recommendation}.
\newblock \bibinfo{journal}{\emph{World Wide Web}} \bibinfo{volume}{27}, \bibinfo{number}{5} (\bibinfo{date}{Aug.} \bibinfo{year}{2024}), \bibinfo{numpages}{31}~pages.
\newblock
\showISSN{1386-145X}
\href{https://doi.org/10.1007/s11280-024-01291-2}{doi:\nolinkurl{10.1007/s11280-024-01291-2}}


\bibitem[Yadav et~al\mbox{.}(2023)]%
        {ties}
\bibfield{author}{\bibinfo{person}{Prateek Yadav}, \bibinfo{person}{Derek Tam}, \bibinfo{person}{Leshem Choshen}, \bibinfo{person}{Colin~A Raffel}, {and} \bibinfo{person}{Mohit Bansal}.} \bibinfo{year}{2023}\natexlab{}.
\newblock \showarticletitle{Ties-merging: Resolving interference when merging models}.
\newblock \bibinfo{journal}{\emph{Advances in Neural Information Processing Systems}}  \bibinfo{volume}{36} (\bibinfo{year}{2023}), \bibinfo{pages}{7093--7115}.
\newblock


\bibitem[Yang et~al\mbox{.}(2024a)]%
        {Survery_ModelMerging_2024}
\bibfield{author}{\bibinfo{person}{Enneng Yang}, \bibinfo{person}{Li Shen}, \bibinfo{person}{Guibing Guo}, \bibinfo{person}{Xingwei Wang}, \bibinfo{person}{Xiaochun Cao}, \bibinfo{person}{Jie Zhang}, {and} \bibinfo{person}{Dacheng Tao}.} \bibinfo{year}{2024}\natexlab{a}.
\newblock \showarticletitle{Model Merging in LLMs, MLLMs, and Beyond: Methods, Theories, Applications and Opportunities}.
\newblock \bibinfo{journal}{\emph{arXiv preprint arXiv:2408.07666}} (\bibinfo{year}{2024}).
\newblock


\bibitem[Yang et~al\mbox{.}(2024b)]%
        {AdaMerging_ICLR_2024}
\bibfield{author}{\bibinfo{person}{Enneng Yang}, \bibinfo{person}{Zhenyi Wang}, \bibinfo{person}{Li Shen}, \bibinfo{person}{Shiwei Liu}, \bibinfo{person}{Guibing Guo}, \bibinfo{person}{Xingwei Wang}, {and} \bibinfo{person}{Dacheng Tao}.} \bibinfo{year}{2024}\natexlab{b}.
\newblock \showarticletitle{AdaMerging: Adaptive Model Merging for Multi-Task Learning}.
\newblock \bibinfo{journal}{\emph{The Twelfth International Conference on Learning Representations}} (\bibinfo{year}{2024}).
\newblock


\bibitem[Yu et~al\mbox{.}(2024)]%
        {DARE}
\bibfield{author}{\bibinfo{person}{Le Yu}, \bibinfo{person}{Bowen Yu}, \bibinfo{person}{Haiyang Yu}, \bibinfo{person}{Fei Huang}, {and} \bibinfo{person}{Yongbin Li}.} \bibinfo{year}{2024}\natexlab{}.
\newblock \showarticletitle{Language Models are Super Mario: Absorbing Abilities from Homologous Models as a Free Lunch}. In \bibinfo{booktitle}{\emph{International Conference on Machine Learning}}. PMLR.
\newblock


\bibitem[Zaman et~al\mbox{.}(2024)]%
        {fuseforget}
\bibfield{author}{\bibinfo{person}{Kerem Zaman}, \bibinfo{person}{Leshem Choshen}, {and} \bibinfo{person}{Shashank Srivastava}.} \bibinfo{year}{2024}\natexlab{}.
\newblock \bibinfo{title}{Fuse to Forget: Bias Reduction and Selective Memorization through Model Fusion}.
\newblock
\showeprint[arxiv]{2311.07682}~[cs.CL]
\urldef\tempurl%
\url{https://arxiv.org/abs/2311.07682}
\showURL{%
\tempurl}


\bibitem[Zang et~al\mbox{.}(2022)]%
        {10.1145/3548455}
\bibfield{author}{\bibinfo{person}{Tianzi Zang}, \bibinfo{person}{Yanmin Zhu}, \bibinfo{person}{Haobing Liu}, \bibinfo{person}{Ruohan Zhang}, {and} \bibinfo{person}{Jiadi Yu}.} \bibinfo{year}{2022}\natexlab{}.
\newblock \showarticletitle{A Survey on Cross-domain Recommendation: Taxonomies, Methods, and Future Directions}.
\newblock \bibinfo{journal}{\emph{ACM Trans. Inf. Syst.}} \bibinfo{volume}{41}, \bibinfo{number}{2}, Article \bibinfo{articleno}{42} (\bibinfo{date}{Dec.} \bibinfo{year}{2022}), \bibinfo{numpages}{39}~pages.
\newblock
\showISSN{1046-8188}
\href{https://doi.org/10.1145/3548455}{doi:\nolinkurl{10.1145/3548455}}


\bibitem[Zhang et~al\mbox{.}(2025)]%
        {zhang2025comprehensivesurveycrossdomainrecommendation}
\bibfield{author}{\bibinfo{person}{Hao Zhang}, \bibinfo{person}{Mingyue Cheng}, \bibinfo{person}{Qi Liu}, \bibinfo{person}{Junzhe Jiang}, \bibinfo{person}{Xianquan Wang}, \bibinfo{person}{Rujiao Zhang}, \bibinfo{person}{Chenyi Lei}, {and} \bibinfo{person}{Enhong Chen}.} \bibinfo{year}{2025}\natexlab{}.
\newblock \bibinfo{title}{A Comprehensive Survey on Cross-Domain Recommendation: Taxonomy, Progress, and Prospects}.
\newblock
\showeprint[arxiv]{2503.14110}~[cs.IR]
\urldef\tempurl%
\url{https://arxiv.org/abs/2503.14110}
\showURL{%
\tempurl}


\bibitem[Zhang et~al\mbox{.}(2023)]%
        {task-arithmetic-lora}
\bibfield{author}{\bibinfo{person}{Jinghan Zhang}, \bibinfo{person}{Shiqi Chen}, \bibinfo{person}{Junteng Liu}, {and} \bibinfo{person}{Junxian He}.} \bibinfo{year}{2023}\natexlab{}.
\newblock \showarticletitle{Composing parameter-efficient modules with arithmetic operations}. In \bibinfo{booktitle}{\emph{Proceedings of the 37th International Conference on Neural Information Processing Systems}} (New Orleans, LA, USA) \emph{(\bibinfo{series}{NIPS '23})}. \bibinfo{publisher}{Curran Associates Inc.}, \bibinfo{address}{Red Hook, NY, USA}, Article \bibinfo{articleno}{552}, \bibinfo{numpages}{22}~pages.
\newblock


\bibitem[Zhao et~al\mbox{.}(2023)]%
        {zhao2023survey}
\bibfield{author}{\bibinfo{person}{Wayne~Xin Zhao}, \bibinfo{person}{Kun Zhou}, \bibinfo{person}{Junyi Li}, \bibinfo{person}{Tianyi Tang}, \bibinfo{person}{Xiaolei Wang}, \bibinfo{person}{Yupeng Hou}, \bibinfo{person}{Yingqian Min}, \bibinfo{person}{Beichen Zhang}, \bibinfo{person}{Junjie Zhang}, \bibinfo{person}{Zican Dong}, {et~al\mbox{.}}} \bibinfo{year}{2023}\natexlab{}.
\newblock \showarticletitle{A survey of large language models}.
\newblock \bibinfo{journal}{\emph{arXiv preprint arXiv:2303.18223}} (\bibinfo{year}{2023}).
\newblock


\bibitem[Zhao et~al\mbox{.}(2024)]%
        {lego}
\bibfield{author}{\bibinfo{person}{Ziyu Zhao}, \bibinfo{person}{Tao Shen}, \bibinfo{person}{Didi Zhu}, \bibinfo{person}{Zexi Li}, \bibinfo{person}{Jing Su}, \bibinfo{person}{Xuwu Wang}, \bibinfo{person}{Kun Kuang}, {and} \bibinfo{person}{Fei Wu}.} \bibinfo{year}{2024}\natexlab{}.
\newblock \showarticletitle{Merging loras like playing lego: Pushing the modularity of lora to extremes through rank-wise clustering}.
\newblock \bibinfo{journal}{\emph{arXiv preprint arXiv:2409.16167}} (\bibinfo{year}{2024}).
\newblock


\bibitem[Zhou et~al\mbox{.}(2022)]%
        {fmlp}
\bibfield{author}{\bibinfo{person}{Kun Zhou}, \bibinfo{person}{Hui Yu}, \bibinfo{person}{Wayne~Xin Zhao}, {and} \bibinfo{person}{Ji-Rong Wen}.} \bibinfo{year}{2022}\natexlab{}.
\newblock \showarticletitle{Filter-enhanced MLP is All You Need for Sequential Recommendation}. In \bibinfo{booktitle}{\emph{Proceedings of the ACM Web Conference 2022}} (Virtual Event, Lyon, France) \emph{(\bibinfo{series}{WWW '22})}. \bibinfo{publisher}{Association for Computing Machinery}, \bibinfo{address}{New York, NY, USA}, \bibinfo{pages}{2388–2399}.
\newblock
\showISBNx{9781450390965}
\href{https://doi.org/10.1145/3485447.3512111}{doi:\nolinkurl{10.1145/3485447.3512111}}


\bibitem[Zhu et~al\mbox{.}(2021)]%
        {ijcai2021p639}
\bibfield{author}{\bibinfo{person}{Feng Zhu}, \bibinfo{person}{Yan Wang}, \bibinfo{person}{Chaochao Chen}, \bibinfo{person}{Jun Zhou}, \bibinfo{person}{Longfei Li}, {and} \bibinfo{person}{Guanfeng Liu}.} \bibinfo{year}{2021}\natexlab{}.
\newblock \showarticletitle{Cross-Domain Recommendation: Challenges, Progress, and Prospects}. In \bibinfo{booktitle}{\emph{Proceedings of the Thirtieth International Joint Conference on Artificial Intelligence, {IJCAI-21}}}, \bibfield{editor}{\bibinfo{person}{Zhi-Hua Zhou}} (Ed.). \bibinfo{publisher}{International Joint Conferences on Artificial Intelligence Organization}, \bibinfo{pages}{4721--4728}.
\newblock
\href{https://doi.org/10.24963/ijcai.2021/639}{doi:\nolinkurl{10.24963/ijcai.2021/639}}
\newblock
\shownote{Survey Track}.


\end{thebibliography}

\appendix
\section{APPENDIX}
\subsection{The Feasibility of LoRA Merging} \label{appendix A.2}
% 证明LoRA自身就是任务向量，操作多LoRA的组合等价与操作多任务的组合，即操作模型的行为
In this subsection, we discuss the feasibility of LoRA merging. Task vector\cite{taskarithmetic} is a concept that describes the change in a model's behavior before and after fine-tuning. Let $\Theta_{pre}\in\mathbb{R}^d$ be the weights of an arbitrary pre-trained model, and $\Theta_{ft}^t\in\mathbb{R}^d$ be the corresponding weights after fine-tuning on task $t$. The task vector $\tau_t$ can be expressed as:
\begin{equation}
    \tau_t=\Theta_{ft}^t-\Theta_{pre}.
\end{equation}

\begin{figure} [htbp]
    \centering
    \subfloat[Task Vector]{\includegraphics[width=0.40\columnwidth]{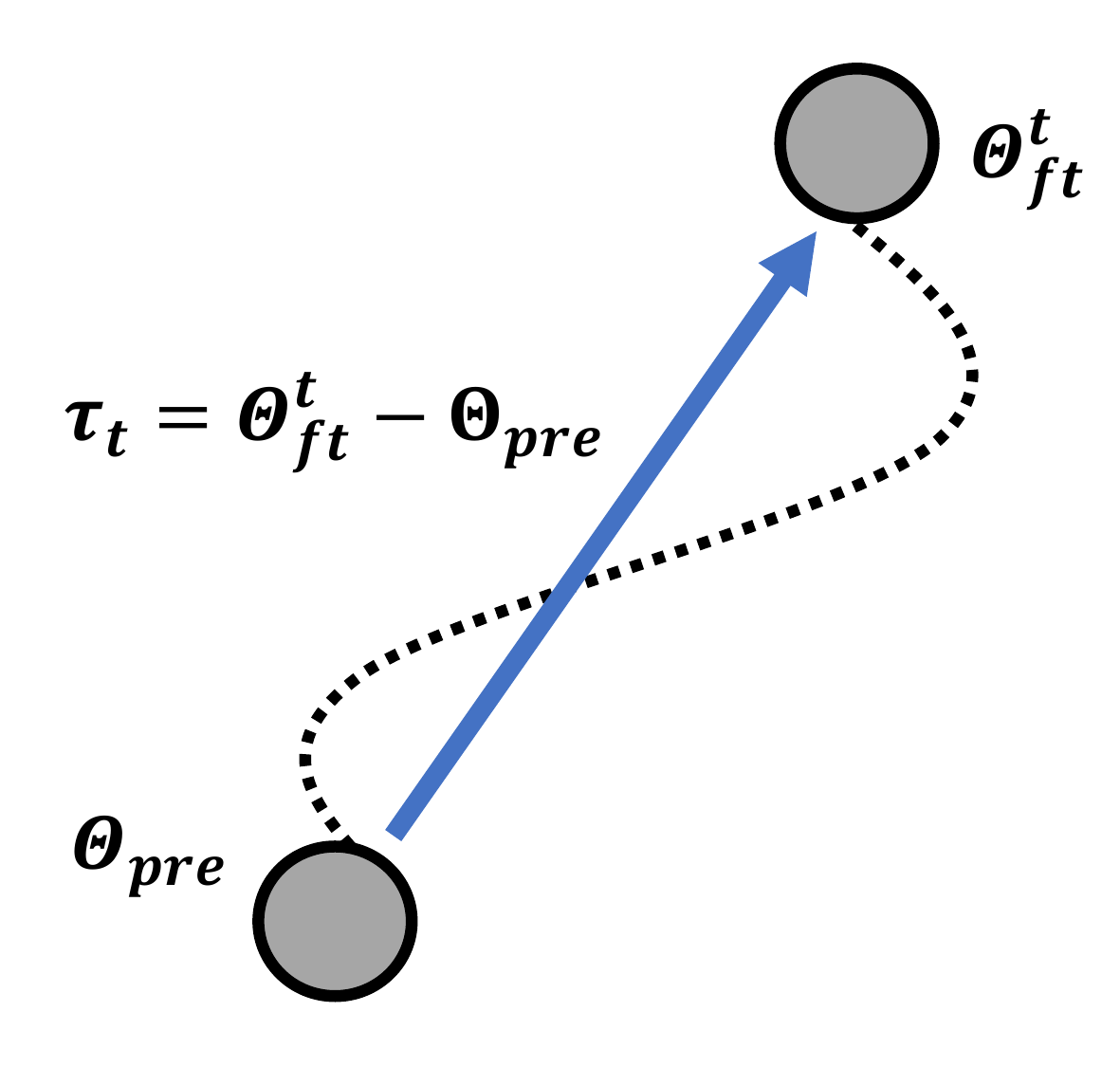}}
    \hfil
    \subfloat[Learning via addition]{\includegraphics[width=0.4\columnwidth]{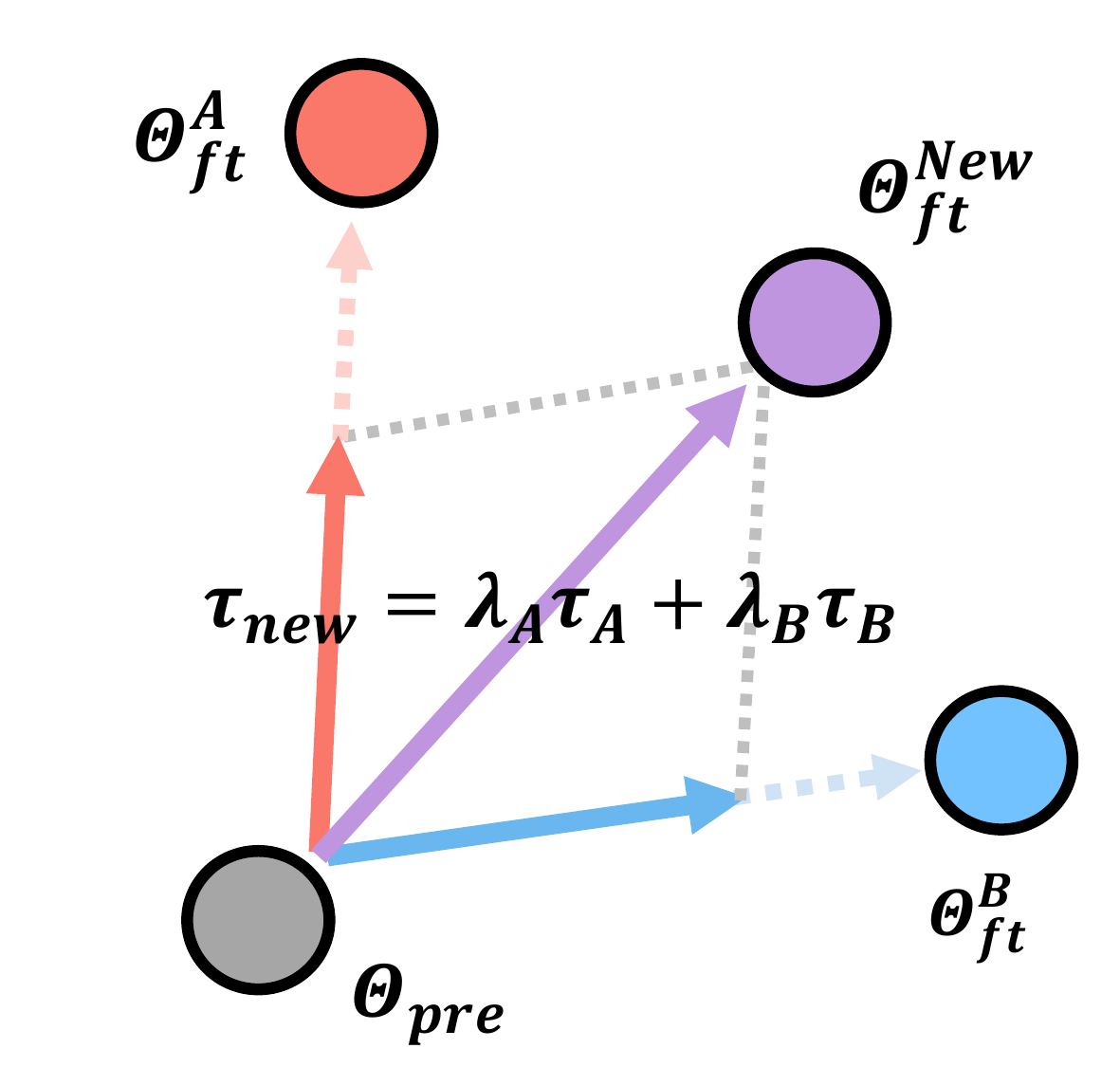}}
    \caption{An illustration of task vector and arithmetic operation. (a) A task vector refers to the difference between the weights of a pre-trained model and its weights after finetuning. (b) Weighted arithmetic operations on a set of task vectors can enhance generalization, consequently boosting model performance.}
    \label{fig:task_vector}
\end{figure}

As shown in Figure \ref{fig:task_vector}, we can manipulate model behavior by performing arithmetic operations on task vectors from different tasks, thereby enhancing the model's performance on a specific target task. For LLMs with LoRA, it is worth noting that the fine-tuned LoRA itself serves as a task vector. Referring to the Equation (\ref{eq2}) in Section \ref{Preliminaries}, a layer in the LLM without LoRA can be equivalently written as:
\begin{equation}
    \boldsymbol{h=Wx+B^0A^0x}, 
\end{equation}
where $\boldsymbol{B^0}$ and $\boldsymbol{A^0}$ are zero matrices. Therefore, after the LLM is loaded with fine-tuned LoRA, the task vector $\tau$ of this layer can be formulated as:
\begin{align}
    \tau&=(\boldsymbol{W+BA})-(\boldsymbol{W+B^0A^0}) \nonumber \\
    &= \boldsymbol{BA}.
\end{align}
From the perspective of the entire model, the task vector is represented as:
\begin{align}
    \tau&=(\Theta+\theta_{LoRA})-\Theta \nonumber \\
    &= \theta_{LoRA}.
\end{align}
Given that performance on a specific task can be enhanced through arithmetic operations on task vectors, the same principle applies to LoRA merging.

\subsection{Dataset} \label{appendix A.3}
The statistics of all datasets are shown in Table \ref{tab:datasets}.

\begin{table}[htbp]
    \centering
    \caption{Statistics of all involved datasets}
    \label{tab:datasets}
    \resizebox{0.45\textwidth}{!}{
    \begin{tabular}{lcccc}
        \toprule
        \textbf{Datasets} & \textbf{\# Users} & \textbf{\# Items} & \textbf{\# Interactions} & \textbf{Density(\%)} \\
        \midrule
        Clothing & 39,387 & 23,033 & 278,677 & 0.0307 \\
        Beauty & 22,363 & 12,101 & 198,502 & 0.0734 \\
        Sports & 35,598 & 18,357 & 296,337 & 0.0453 \\
        Food & 14,681 &  8,713 & 151,254 & 0.1182 \\
        Toys & 19,412 & 11,924 & 167,597 & 0.0724 \\
        Movielens-1M & 6,040 &  6,883 &  1,000,209 &  2.4059 \\
        \bottomrule
    \end{tabular}
    }
\end{table}
\subsection{Baseline} \label{appendix A.4}
The following provides an overview of all baselines. \textbf{GRU4Rec}~\cite{gru4rec} is a seminal method that uses RNNs to model user action sequences for session-based recommendation. \textbf{SASRec}~\cite{sasrec}  employs self-attention mechanisms to model long-term dependencies in user interaction history. \textbf{BERT4Rec} \cite{bert4rec} adapts the bidirectional transformer architecture from BERT to sequential recommendation. \textbf{FMLP-Rec}~\cite{fmlp} is an all-MLP model with learnable filters for sequential recommendation tasks. \textbf{MCRPL}~\cite{mcrpl} proposes a two-stage prompting-based paradigm for challenges such as the absence of overlapping information and distribution discrepancy between different domains.
\textbf{VQ-Rec}~\cite{hou2023vqrec} and \textbf{UniSRec}~\cite{unisrec} employ contrastive pre-training on language models to learn domain-agnostic representations that facilitate knowledge transfer without requiring explicit overlaps.
\textbf{RecFormer}~\cite{recformer} models user preferences and item features using the LongFormer backbone, transforming sequential recommendation into a task of predicting the next item as if predicting the next sentence, by converting item attributes into a sentence format.
\textbf{Qwen2-7B} is a well-known open-source LLM. We use the zero-shot version. 
\textbf{TALLRec}~\cite{tallrec} learns the recommendation task based on prompts consisting solely of text and fine-tunes the LLMs using the LoRA.
textbf{LLM-Rec}~\cite{one-model-for-all} 
adopt descriptive information of users’ mixed sequence from multidomain to build universal representation via LLM.
\textbf{Weight Average} directly averages the weights of multiple single-domain models.
\textbf{AdaMerging}~\cite{AdaMerging_ICLR_2024} is an adaptive model merging technique that automatically learns optimal merging coefficients for multi-task learning.
\textbf{LoRA-LEGO} \cite{lego} is a modular LoRA merging framework that treats each rank in LoRA as a Minimal Semantic Unit (MSU) and merges multiple LoRAs through rank-wise clustering, enabling flexible disassembly and reassembly of LoRA modules while mitigating parameter interference.
\textbf{Ties-Merging}~\cite{ties} involves a three-step process that includes reducing parameter redundancy, eliminating sign conflicts between parameters, and finally merging them.
\textbf{Target-Domain Only} denotes the model that is solely trained with target domain data.
\textbf{All Data Merging} denotes the model that is trained with all source-domain and target-domain data.

\subsection{Experiment Implementation Details} \label{appendix A.5}
We use Qwen2-7B as the LLM backbone for WeaveRec. For parameter-efficient finetuning (PEFT) methods, we adopt low-rank adaption (LoRA) with LoRA rank as 16, LoRA alpha as 32, and LoRA dropout as 0.05 to get different LoRA adapters. For Hybrid LoRAs, due to computational time constraints and other factors, we sampled 40,000 training samples from each of the two involved domains. For each domain, this sample size represents a proportion ranging between 60\% and 100\% of the full dataset. A domain with a smaller dataset size consequently exhibits a higher sampling proportion in our approach. The learning rate is set to 2e-4 and the batch size is set to 128. In order to reduce GPU memory usage, we used gradient checkpointing techniques. And we use the VLLM inference acceleration framework to perform inference and then evaluate the results.

To ensure a fair comparison, the experimental settings are standardized as follows. For single-domain sequential recommendation methods (GRU4Rec, SASRec, BERT4Rec, and FMLP-Rec), the learning rate is set to 0.001, and the Adam optimizer is employed. The batch size is set to 256 and the embedding dimension is set to 64. With respect to cross-
domain sequential recommendation methods, VQ-Rec and UnisRec utilize a BERT for text processing. VQ-Rec, UnisRec, and RecFormer are fine-tuned and then tested using our target domain data, based on the pre-trained parameters provided by their respective authors. MCRPL first undergoes pre-training using the target domain and all source domains, and then is fine-tuned and tested on each of its respective domains. Regarding LLM-based Recommendation methods, Qwen2-7B is directly zero-shot tested on the target domain. TALLRec is fine-tuned on data from all target and source domains, and subsequently evaluated on each domain individually, whose backbone is Llama2-7B\footnote{\url{https://huggingface.co/meta-llama/Llama-2-7b-chat-hf}}. LLM-REC, similar to TALLRec, is also fine-tuned on all domain data and then tested individually on each domain. According to the original text, its backbone is BERT. For Model Merging Methods, they all integrate the target-domain model and all source-domain models according to their respective methods to form a new, single model, which is then tested individually on each domain. Among these, AdaMerging randomly select 50 unlabeled test data samples from each domain and combined them, which is used for Test-Time Adaptation to learn the fusion weights. For Our ablation counterparts, the LoRA and other experimental parameters used are consistent with those of WeavRec. All experiments are conducted on 8 NVIDIA GeForce RTX 4090 (24GB) GPUs.

\subsection{\textbf{More In-Depth Analysis}} \label{more-analysis}
\subsubsection{\textbf{Why one source domain per branch?}}
In this section, we explore why each branch of \textit{WeaveRec} contains at most one source domain rather than combining multiple source domains into one branch. To investigate this, we conducted controlled experiments, fixing \textit{WeaveRec} to two branches. As shown in Figure \ref{indepth-1}, "0 Source Domain" signifies a degeneration to the target-domain LoRA, while larger values indicate mixing all source domains with the target domain in one branch.
Performance is optimal when only one source domain is mixed within a single branch. A substantial decline in model performance was observed when multiple source domains were mixed in one branch. When data from one source domain are mixed with that of the target domain to form the second branch, \textit{WeaveRec}'s performance surpasses that of target-domain LoRA, indicating the alleviation of performance degradation. However, when multiple source domains are mixed with the target domain to form the second branch, performance degradation of varying degrees occurs, suggesting the problem likely caused by potential gradient interference and other factors inherent in multi-task learning. This explains why \textit{WeaveRec} allocates source domains to separate branches before merging.

\subsubsection{\textbf{Hybrid model outperforms source-domain model.}}
\begin{figure}[htbp]
    \centering
    \subfloat[]{\includegraphics[width=0.46\columnwidth]{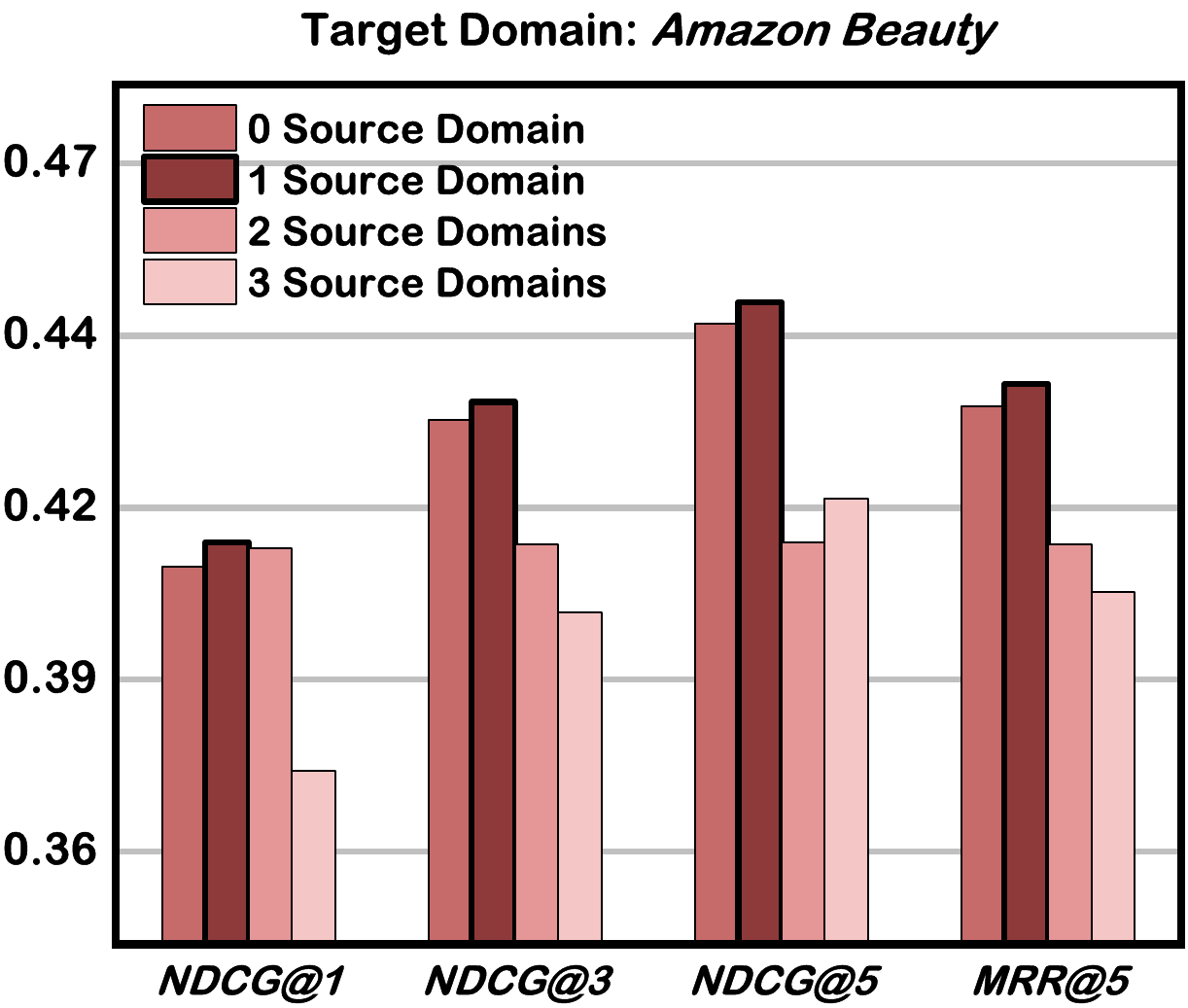}\label{indepth-1}}
    \hfil
    \subfloat[]{\includegraphics[width=0.49\columnwidth]{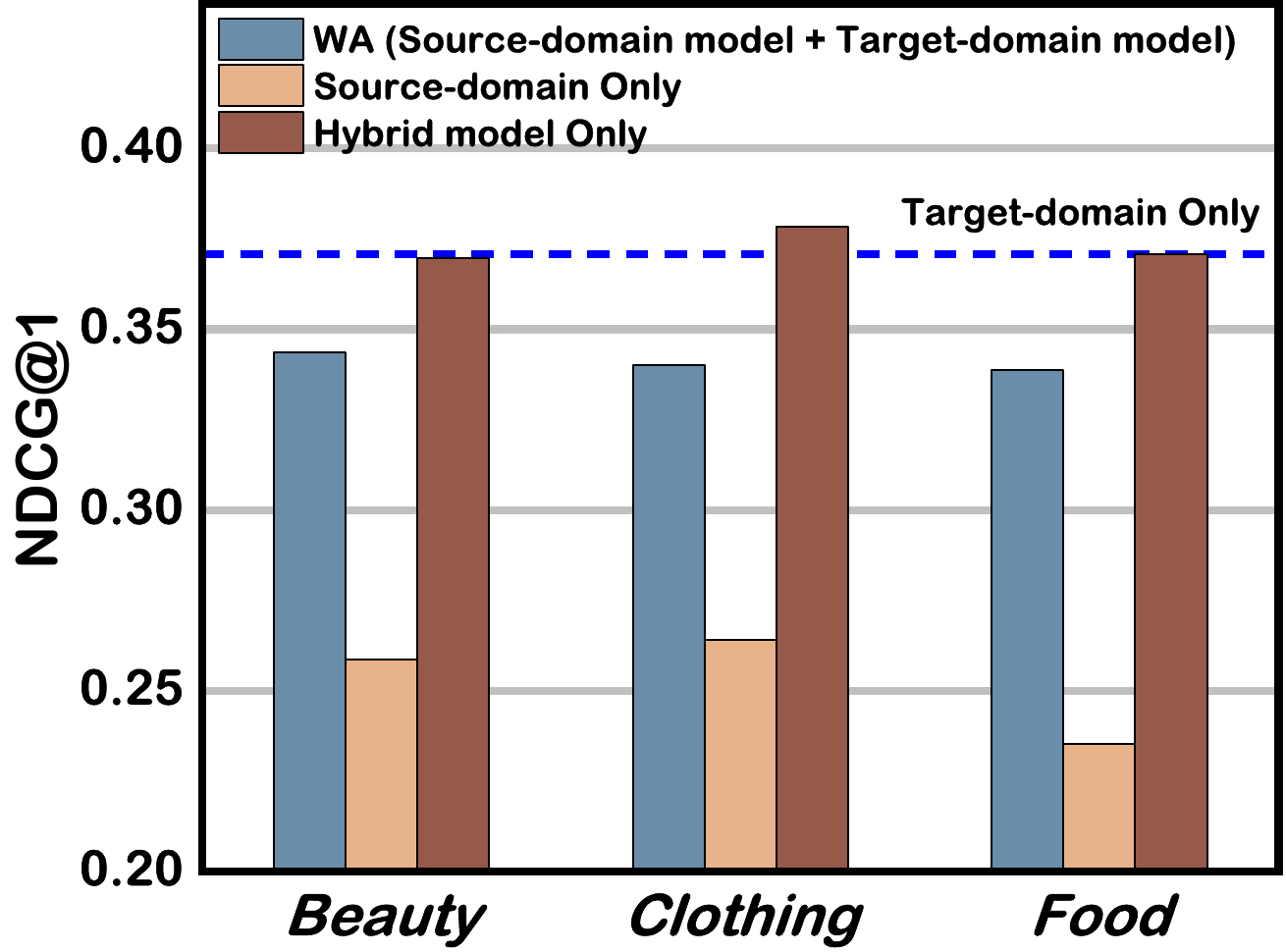}\label{sd-hybrid}}
    \caption{(a) \textit{WeaveRec}'s performance comparison under fixed two-branch conditions. The three source domains are, in order, Amazon Food, Sports, and Clothing. (b) Comparison of performance on the fixed target domain Sports under different source domain conditions.} 
    %\vspace{-0.4cm}
    \label{mmm} 
\end{figure}

Here, we expand upon Figure \ref{fig:sub_b}. As shown in Figure \ref{sd-hybrid}, for the same target domain but with different source domains, we obtain their respective hybrid models. The different hybrid models all perform comparably to the target-domain model on the target domain, and they all perform much better than their corresponding source-domain models. As a key component of WeaveRec, the hybrid model demonstrates its superiority and rationality over the source-domain model in Figure \ref{sd-hybrid}, confirming the findings in the Section \ref{analysis}.

\section{Related Work}
\noindent$\bullet$ \quad\textbf{Cross-Domain Sequential Recommendation.} Cross-domain recommendation~\cite{10.1145/3548455,ijcai2021p639,zhang2025comprehensivesurveycrossdomainrecommendation} seeks to improve user preference modeling in target domain by transferring knowledge from multiple source domains. Existing CDSR methodologies can be divided into two primary categories based on their representation strategies: 1) ID-based approaches~\cite{10.1109/TKDE.2022.3185101,10.1609/aaai.v38i8.28723,10.1145/3539597.3570366,10.1109/TKDE.2024.3511602}, which employ collaborative filtering models to learn domain-specific embeddings, which are subsequently aligned through overlapping users or items via techniques such as mapping functions or shared latent spaces. While effective when overlap exists, these methods face severe scalability constraints due to their dependency on cross-domain overlaps, which are often sparse or unavailable in practice. To overcome this limitation, PLCR\cite{PLCR} is an automated prompting-based recommendation framework for non-overlapping scenarios. MCRPL\cite{mcrpl} proposes a two-stage prompting-based paradigm for challenges such as the absence of overlapping information and distribution discrepancy between different domains. 2) Transferable approaches~\cite{unisrec,recformer,hou2023vqrec} address this limitation by utilizing content-based representations, particularly textual descriptions, to encode items in a unified semantic space. Notable examples include VQ-Rec~\cite{hou2023vqrec} and UniSRec~\cite{unisrec}, which employ contrastive pre-training on language models to learn domain-agnostic representations that facilitate knowledge transfer without requiring explicit overlaps.
Recently, these methods have been further advanced through the incorporation of LLMs' multi-domain integration capabilities~\cite{m6rec,one-model-for-all,ecellm}. However, transferable-based works primarily focus on improving the model's overall capabilities across multiple domains, while overlooking the negative transfer phenomenon. We find that this may lead to performance degradation in the target domain.

\noindent$\bullet$ \quad\textbf{LLM-Based Recommendation.}
The emergence of LLMs has catalyzed a paradigm shift in recommender systems, giving rise to LLM-based recommendation approaches that directly leverage LLMs as recommendation engines~\cite{10.1145/3678004,10.1007/s11280-024-01291-2}.
Early studies~\cite{10.1145/3604915.3610646,10.1145/3604915.3608845,wang2023rethinking} explore LLM's zero-shot/few-shot potential via in-context learning. However the gap between LLMs' pretraining on general text and recommendation-specific needs leads to suboptimal performance. To address this, constructing recommendation data into text-based instruction fine-tuning datasets and supervised fine-tuning LLMs has been validated to be effective~\cite{tallrec}.
The rapid rise of LLMs is shifting recommender systems from task-specific designs to unified, general-purpose models capable of handling diverse domains and tasks~\cite{LLM4CDR,one-model-for-all,p5,m6rec,ecellm}. This "one model for all" paradigm capitalizes on LLMs' capacity to encode heterogeneous data sources and perform cross-domain inference through prompt-driven frameworks. Representative works in this direction include P5~\cite{p5}, which designs unified prompts to integrate five distinct recommendation tasks within a text-to-text framework; M6-rec~\cite{m6rec}, which develops a foundation model supporting open-ended domains and tasks in industrial settings; LLM-Rec~\cite{one-model-for-all,bao2025heterogeneoususermodelingllmbased}, which explores language models' capabilities in modeling multi-domain user behavior. These approaches collectively challenge the traditional single-domain, single-task recommendation paradigm and demonstrate significant practical value. Nevertheless, they are primarily focus on the "data merging" strategies. It may suffer from fundamental scalability and flexibility limitations: the addition of a new domain or task necessitates model retraining from scratch, resulting in prohibitive computational costs.

\noindent$\bullet$ \quad\textbf{Model Merging.}
Traditional ensembling~\cite{dong2020survey} improves performance by averaging predictions from multiple models, but this approach 
comes with the downside of increased inference costs and becomes fundamentally impractical for large language models since their text outputs cannot be meaningfully averaged or merged. Model merging~\cite{Survery_ModelMerging_2024} offers a viable alternative by combining model parameters in weight space rather than attempting to merge outputs. Model merging is rooted in the theoretical foundation of mode connectivity~\cite{1996Weight,10.5555/3327546.3327556,frankle2020linear}, the principle that models fine-tuned from the same pre-trained checkpoint often reside in connected regions of the loss landscape, enabling meaningful parameter interpolation without significant performance degradation. 
Early work~\cite{wortsman2022model} demonstrated the effectiveness of simple arithmetic averaging of corresponding parameters across models. This was subsequently extended by task arithmetic~\cite{taskarithmetic,task-arithmetic-lora} approaches that treat model parameter differences as vectors, enabling mathematical operations like addition, subtraction, and scaling to combine or remove specific capabilities. More sophisticated methods~\cite{lego,AdaMerging_ICLR_2024,ext-sub,ties} have emerged to improve performance across multiple tasks by reducing conflicts between models and adjusting merging weights.

\end{document}